\documentclass{article}
\usepackage{ijcai25}

\usepackage{times}
\usepackage{soul}
\usepackage{url}
\usepackage[hidelinks]{hyperref}
\usepackage[utf8]{inputenc}
\usepackage[small]{caption}
\usepackage{graphicx}
\usepackage{amsmath}
\usepackage{amsthm}
\usepackage{booktabs}
\usepackage{algorithm}
\usepackage{algorithmic}
\usepackage[switch]{lineno}

\usepackage{caption}
\usepackage{subcaption}
\usepackage{multirow}
\newtheorem{definition}{Definition}

\usepackage{pifont}
\newcommand{\cmark}{\ding{51}}%
\newcommand{\xmark}{\ding{55}}%
\usepackage[normalem]{ulem}
\useunder{\uline}{\ul}{}
\usepackage{amssymb}
\usepackage{wrapfig}

\newtheorem{theorem}{Theorem}

\title{Heterogeneous Temporal Hypergraph Neural Network}

\author{
Huan Liu$^{1,2}$\and
Pengfei Jiao$^{1,2}$\and
Mengzhou Gao$^{1,2}$\footnote{Corresponding authors.}\and
Chaochao Chen$^3$\And
Di Jin$^4$\footnotemark[1]\\
\affiliations
$^1$School of Cyberspace, Hangzhou Dianzi University, Hangzhou, China\\
$^2$Data Security Governance Zhejiang Engineering Research Center, Hangzhou, China\\
$^3$College of Computer Science and Technology, Zhejiang University, Hangzhou, China\\
$^4$College of Intelligence and Computing, Tianjin University, Tianjin, China\\
\emails
\{huanliu, pjiao, mzgao\}@hdu.edu.cn,
zjuccc@zju.edu.cn,
jindi@tju.edu.cn
}

\begin{document}

\maketitle

\begin{abstract}
    Graph representation learning (GRL) has emerged as an effective technique for modeling graph-structured data. When modeling heterogeneity and dynamics in real-world complex networks, GRL methods designed for complex heterogeneous temporal graphs (HTGs) have been proposed and have achieved successful applications in various fields. However, most existing GRL methods mainly focus on preserving the low-order topology information while ignoring higher-order group interaction relationships, which are more consistent with real-world networks. In addition, most existing hypergraph methods can only model static homogeneous graphs, limiting their ability to model high-order interactions in HTGs. Therefore, to simultaneously enable the GRL model to capture high-order interaction relationships in HTGs, we first propose a formal definition of heterogeneous temporal hypergraphs and $P$-uniform heterogeneous hyperedge construction algorithm that does not rely on additional information. Then, a novel \underline{H}eterogeneous \underline{T}emporal \underline{H}yper\underline{G}raph \underline{N}eural network (HTHGN), is proposed to fully capture higher-order interactions in HTGs. HTHGN contains a hierarchical attention mechanism module that simultaneously performs temporal message-passing between heterogeneous nodes and hyperedges to capture rich semantics in a wider receptive field brought by hyperedges. Furthermore, HTHGN performs contrastive learning by maximizing the consistency between low-order correlated heterogeneous node pairs on HTG to avoid the low-order structural ambiguity issue. Detailed experimental results on three real-world HTG datasets verify the effectiveness of the proposed HTHGN for modeling high-order interactions in HTGs and demonstrate significant performance improvements.
\end{abstract}

\section{Introduction}
\label{sec:intro}

Graph representation learning (GRL) has emerged as an effective technique for learning real-world graph-structured data and has been widely applied in various fields~\cite{9416834,gao2023hierarchical,10.1145/3527662}. In the major research of GRL, graph neural networks (GNNs) as encoders have gained universal effectiveness due to their powerful message-passing mechanism and fitting ability~\cite{kipf2017semisupervised,10.5555/3294771.3294869}. However, most methods assume that the network is homogeneous and static and only includes pairwise relationships, which often contradicts real-world systems~\cite{10.1145/3605776,10.1145/3483595,9780569}. For example, in an academic network that includes multiple node types such as \textit{author}, \textit{paper} and \textit{venue}, as well as evolving \textit{co-author} and \textit{co-citations} relationships among multiple authors and papers over time. This network contains heterogeneity and dynamics in not low-order but group interactions, which are too complex to be described by simple pairwise graphs~\cite{10.1145/3605776}. 

Considering the successful applications of GRL methods of heterogeneous graphs, dynamic graphs, and hypergraphs, we propose a formal definition of \textit{heterogeneous temporal hypergraphs} (HTHGs) as a modeling tool to comprehensively describe high-order relationships in heterogeneous temporal graphs (HTGs). Specifically, HTHGs refer to hypergraphs that contain high-order relationships among three or more multi-type entities, and these entities and relationships could increase or delete over time. Since HTHGs involve multiple types of nodes and interaction patterns, effectively modeling the high-order correlations and semantic information inherent in HTHGs is crucial for representation learning.

However, existing GRL and GNN research usually one-sidedly simplifies HTHG in different aspects, thus losing its information integrity and seriously affecting performance. For example, on the one hand, modeling group interactions as hyperedges and performing representation learning through hypergraph GNNs has become a dominant paradigm and has achieved remarkable results~\cite{ijcai2021p353,9795251}. However, these methods usually only focus on homogeneous hypergraphs with static structures, failing to model the heterogeneity and dynamics in HTGs. On the other hand, for heterogeneity and dynamics, GNNs designed for heterogeneous graphs~\cite{9780569,9300240,9428609} and dynamic graphs~\cite{10.1145/3483595,ZHANG2023109486,9714053} have been proposed respectively and achieved encouraging results. However, this method usually performs representation learning on pairwise networks and cannot model both low-order and high-order dynamic relationships between multiple heterogeneous nodes.

Although effectively modeling semantic information, temporal dependence, and group interactions has demonstrated significant performance improvements in practice, uniformly modeling HTHGs still has the following challenges: \textbf{1) How to model group interactions without relying on expert knowledge?} Additional information or predefined structures are required to model group interactions, but their effectiveness depends on prior knowledge and is difficult to generalize. \textbf{2) How to model both low-order and high-order information on HTHGs?} The message-passing mechanism of the GNNs and cluster/star expansion are often used to model high-order interaction, but they have computational problems~\cite{10.1145/3605776} and cannot preserve both low-order and high-order relationships simultaneously. \textbf{3) How to perform message-passing on HTHGs?} To preserve temporal structural and semantic information in latent representation space, enhancing communication between heterogeneous nodes and hyperedges is necessary and challenging.

In this paper, to uniformly model the complex blend of heterogeneity, dynamics, and group interactions, we first provide a formal definition of HTHGs, which describes multi-scale group interaction relationships containing dynamics and heterogeneity.
Furthermore, to universally model high-order interactions in HTHGs, we formally define two general heterogeneous hyperedges, $k$-hop, and $k$-ring, and $P$-uniform hyperedges based on high-order neighborhood sampling, which do not rely on predefined structures.
Then, we propose a novel contrastive heterogeneous temporal hypergraph neural network called HTHGN to capture the high-order dynamic semantics contained in HTHGs. Specifically, HTHGN contains a hierarchical attention mechanism that simultaneously performs cross-temporal message passing between heterogeneous nodes and hyperedges to capture rich semantic information in a wider receptive field brought by hyperedges.
Finally, to avoid the problem of low-order structural ambiguity, a heterogeneous low-order structure-preserving contrastive learning objective function is used to optimize the overall HTHGN. Detailed experimental results on $3$ real-world datasets demonstrate that group interaction significantly gains representation learning and verifies the effectiveness of the proposed contrastive learning method HTHGN.

In summary, the contributions of this paper are as follows: 
\begin{itemize}
    \item We study the complex properties prevalent in real-world complex networks. To the best of our knowledge, we are the first to define HTHGs, which are used to model complex networks containing dynamics, heterogeneity, and group interactions. 
    \item We propose a general hyperedges construction algorithm to model high-order semantic information without relying on additional information and prior knowledge.
    \item We propose a novel contrastive heterogeneous temporal hypergraph neural network, HTHGN, which simultaneously models low-order and high-order interactions, and extensive experimental results on 3 real-world datasets verify its superior performance.
\end{itemize}

\section{Background and Preliminaries}
\label{sec:pre}

\begin{figure*}[!htb]
    \centering
    \includegraphics[width=\linewidth]{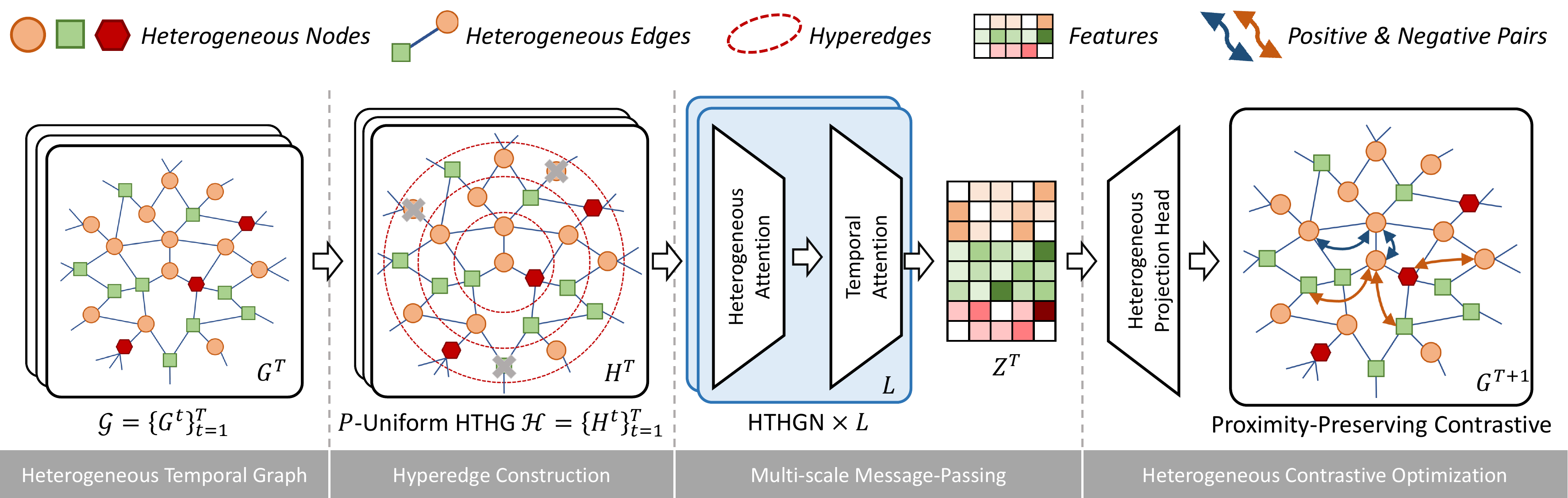}
    \caption{Overall architecture of the proposed HTHGN model.}
    \label{fig:overall}
\end{figure*}

The overall architecture of our proposed HTHGN model is shown in Figure~\ref{fig:overall}. We design a pipeline to learn node representations, which can capture both low-order and high-order information within a HTG.
The basic definitions in this paper are shown below:

\begin{definition}[\textbf{Heterogeneous Graph}]
A \textbf{Heterogeneous Graph} can be defined as $G=(V, E, X)$, where $V$ and $E$ denote the node set and the edge set, respectively; $X\in\mathbb{R}^{|V|\times D}$ is the $D$-dimensional attribute matrix of nodes. Each node $v \in V$ and link $e \in E$ is associated with their mapping functions $\phi(v): V \rightarrow \mathcal{A}$ and $\psi(e): E \rightarrow \mathcal{R}$, where $\mathcal{A}$ and $\mathcal{R}$ denote the node types and link types, and $|\mathcal{A}|+|\mathcal{R}| > 2$ due to heterogeneity.
\end{definition}

\begin{definition}[\textbf{Heterogeneous Temporal Graph}]
A \textbf{Heterogeneous Temporal Graph} is a list of observed heterogeneous snapshots $\mathcal{G} = \left\{G^{1}, G^{2},\ldots, G^{T} \right\}$ ordered by timestamps, where $T$ is the size of time window and $G^{t}=(V^{t}, E^{t}, X^t)$ represents the $t$-th snapshot. The node set $V^{t}$ and edge set $E^{t}$ can differ between snapshots, representing dynamic addition and removal of nodes and edges. 
\end{definition}

\begin{definition}[\textbf{Link Prediction}]
\label{def:linkpred}
Given a heterogeneous temporal graph $\mathcal{G} = \left\{G^{t}\right\}_{t=1}^{T}$ and the learned node representations $Z \in \mathbb{R}^{|V^{t}|\times d}$, the \textbf{link prediction} is the problem of predicting the probability $p\left((i,j)\in E^\tau \mid z_i, z_j\right)$ where $\tau > T$. Besides, the \textbf{\textit{new} link prediction} predicts the probability $p\left((i,j)\in E^\tau \mid z_i, z_j, (i,j)\notin E^T \right)$ where $\tau >T$.
\end{definition}

\section{Heterogeneous Temporal Hypergraph Contrastive Learning: HTHGN}
\label{sec:method}

\subsection{Hypergraph Construction}
Given a HTG $\mathcal{G} = \{G^t\}_{t=1}^T$, the hypergraph construction module is leveraged to construct heterogeneous collective relations based on the heterogeneous snapshots. For this purpose, it is necessary to employ a graph structure capable of modeling interactions that encompass both multiple node types and collective behavior. Formally, we define as follows:

\begin{definition}[\textbf{Heterogeneous hypergraph}]
A \textbf{heterogeneous hypergraph} can be defined as $H=(V, \mathcal{E})$, where $V$ and $\mathcal{E}$ denote the node set and the hyperedges set, respectively; each hyperedge $e \in \mathcal{E}$ is a subset of $V$, i.e., $e \subseteq V$,
and each \textit{node} $v \in V$ is contained in at least one hyperedge $e \in E$, i.e., $V = \bigcup_{e \in \mathcal{E}} e$. Each node $v \in V$ and link $e \in \mathcal{E}$ is associated with their mapping functions $\phi_h(v): V \rightarrow \mathcal{A}_h$ and $\psi_h(e): E \rightarrow \mathcal{R}_h$, where $\mathcal{A}_h$ and $\mathcal{R}_h$ denote the node types and hyperedge types, and $|\mathcal{A}_h|+|\mathcal{R}_h| > 2$ due to heterogeneity.
When $|\mathcal{A}_h|+|\mathcal{R}_h| = 2$, the heterogeneous hypergraph degenerates into a homogeneous hypergraph, represented as $H^-$.
\end{definition}

Due to heterogeneous hyperedges, $H$ has a higher level of expressive power that can encompass many entity types between subtle relationships. However, hyperedges are typically defined by original data and predefined structures, such as meta-paths, which often introduce noise due to insufficient modeling and rely on specific scenarios, limiting their generalizability. Consequently, we propose heterogeneous network-structure-based hyperedge construction methods for constructing hyperedges, i.e., $k$-hop and $k$-ring heterogeneous hyperedge.

\begin{definition}[\textbf{$k$-hop heterogeneous hyperedge}]
A \textbf{$k$-hop heterogeneous hyperedge} around a node $v \in V$ is defined as the set of all nodes that are \textbf{within} cumulative topological distance $k$ edges from the node $v$, regardless of the types of edges and nodes. We denote the set of nodes in the receptive field of $v$ by $e(v)_{k\text{-hop}}$, which is defined recursively as:
\begin{equation}
    \begin{aligned}
    e(v)_{k\text{-hop}} &:= e(v)_{(k-1)\text{-hop}} \\  
    &\cup\{u\in V \mid (v, u)\in E \land u\in \mathcal{N}_v^{k-1}\},
\end{aligned}
\end{equation}
where $\mathcal{N}_v^{k}$ represents the $k$-hop neighborhoods set of node $v$ and $e(v)_{1\text{-hop}} := \mathcal{N}_v$.
\end{definition}

\begin{figure}[!htb]
    \centering
    \includegraphics[width=0.6\linewidth]{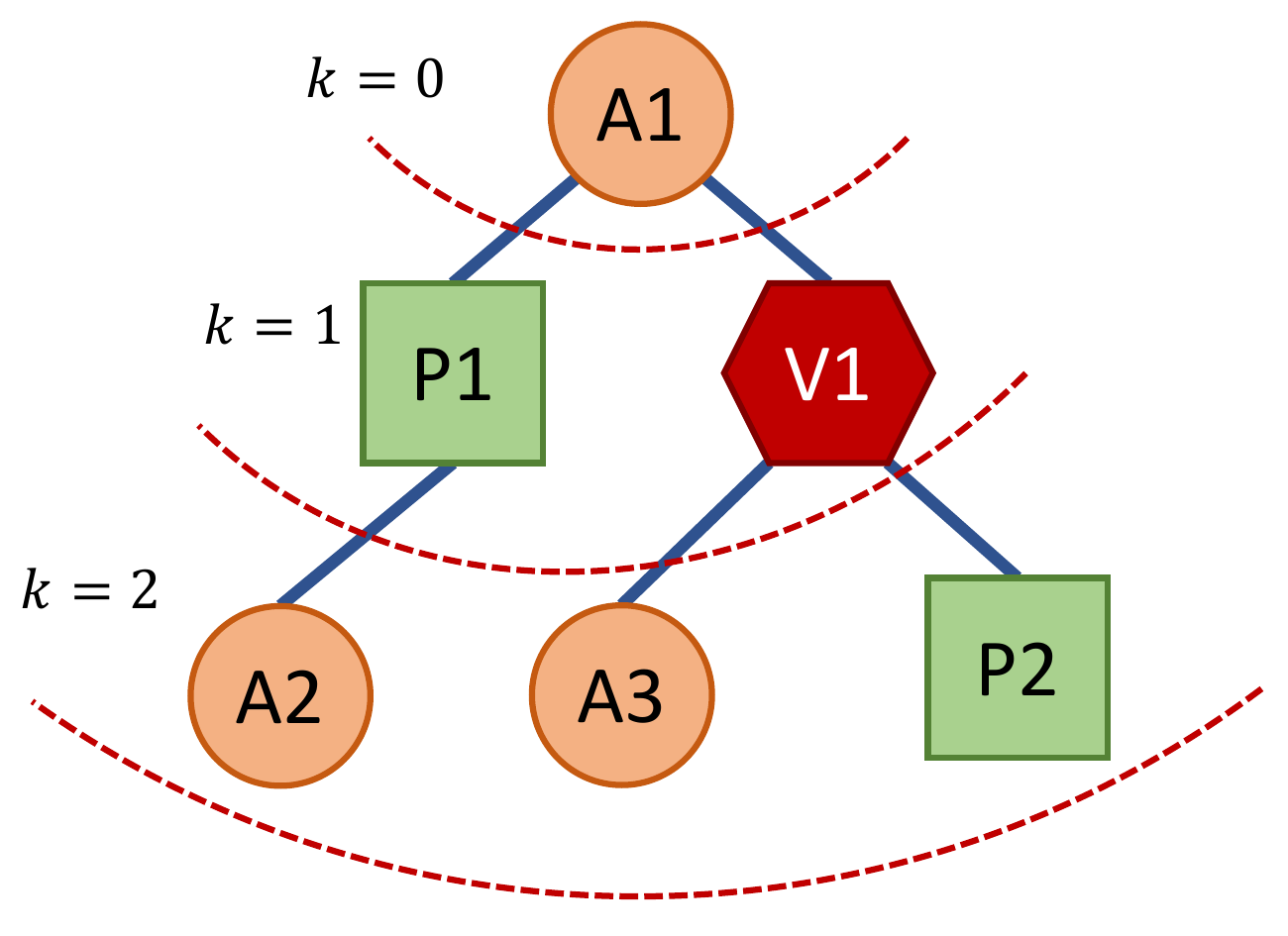}
    \caption{$k$-hop heterogeneous hyperedge toy example.}
    \label{fig:khoptoy}
\end{figure}

This approach to constructing hyperedges based on $k$-hop connectivity implicitly integrates multifaceted semantic layers and relationships between entities, offering a nuanced understanding of their interactions. As illustrated in Figure~\ref{fig:khoptoy}, which has \textit{Author(A)}, \textit{Paper(P)} and \textit{Venue(V)}. when $k=1$, the $1$-hop hyperedge of node \textit{A1} represents all direct neighbors, that is, $e(A1)_{1\text{-hop}}=\{P1, V1\}$. When $k=2$, the $2$-hop hyperedge encompasses both $1$-hop and $2$-hop neighbors, thus, $e(A1)_{2\text{-hop}}=\{P1, V1, A2, A3, P2\}$. This encapsulates rich semantic information that can be recognized as multiple meta-paths, such as \textit{AP}, \textit{AV}, \textit{APA}, \textit{AVA}, \textit{AVP}, \textit{etc}.

\begin{definition}[\textbf{$k$-ring heterogeneous hyperedge}]
A \textbf{$k$-ring heterogeneous hyperedge} around a node $v\in V$ is defined as the set of all heterogeneous nodes that reside \textbf{exactly} a topological distance of $k$ edges from $v$, where the types of nodes and edges along the path are not necessarily the same, denoted as $e(v)_{k\text{-ring}}$, which is defined recursively as:
\begin{equation}
    \begin{aligned}
    e(v)_{1\text{-ring}} &:= \mathcal{N}_v, \\
    e(v)_{k\text{-ring}} &:= \{u\in V \mid (v, u)\in E \land u\in \mathcal{N}_v^{k-1}\}.
\end{aligned}
\end{equation}
    
\end{definition}

Different from $k$-hop heterogeneous hyperedge, the $k$-ring focuses on the heterogeneous nodes at a specific distance, which helps understanding group interactions at that exact distance. Given the substantial augmentation in the number of nodes and edges typically resulting from the $k$-hop/$k$-ring expansion in hypergraphs, we propose the concept of a $P$-uniform heterogeneous hypergraph. 

\begin{definition}[\textbf{$P$-uniform heterogeneous hypergraph}]
\label{def:puniform}
Given $k \in \mathbb{N}$ and $P \in \mathbb{N}$, a \textbf{$P$-uniform hypergraph}, denoted as $H_P=(V, \mathcal{E}_{k})$, is a heterogeneous hypergraph that every $k$-hop/$k$-ring hyperedge $e \in \mathcal{E}_k$ connects exactly $P$ nodes from $V$. That is:
    \begin{equation}
        H_P=(V, \mathcal{E}_{k}) \text{ is } P\text{-uniform} \Leftrightarrow \forall e\in \mathcal{E}_k, |e| = P,
    \end{equation}
where $|e|$ denotes the cardinality, i.e., the number of heterogeneous nodes of the hyperedge $e$.
\end{definition}
This $P$-uniform heterogeneous hypergraph addresses the computational and complexity challenges inherent to hypergraph expansions by introducing a uniform sampling of hyperedges, exhibiting invariance in heterogeneous hyperedge cardinality.

\begin{theorem}[\textbf{Scalability of $k$-hop/$k$-ring structures in $P$-uniform heterogeneous hypergraph}]
\label{theorem:scala}
For a $P$-uniform heterogeneous hypergraph $H_P = (V, \mathcal{E}_k)$, let $|V| \rightarrow \infty$ while maintaining $|e| = P, \forall e \in \mathcal{E}_k$. It follows that the number of $k$-hop/$k$-rings structures grows polynomially with the size of $V$, assuming a constant average degree.
\end{theorem}
The $P$-uniform heterogeneous hypergraph condenses the hypergraph by uniformly sampling the incorporating nodes and thus balanced representation of the underlying heterogeneous relationships while mitigating the exponential increase in graph components. Then, we construct a HTHG based on the heterogeneous hypergraph snapshots.

\begin{definition}[\textbf{Heterogeneous Temporal Hypergraph}]
A Heterogeneous Temporal Hypergraph is denoted as $\mathcal{H}=\{H^t\}_{t=1}^T$, where each $H^t = (V^t, \mathcal{E}^t)$ is a snapshot of the hypergraph at time $t$. Each snapshot consists of a node set $V^t$, hyperedges set $\mathcal{E}^t$, type-assignment functions for nodes $\phi_h: V^t \rightarrow \mathcal{A}_h$ and for hyperedges $\psi_h: \mathcal{E}^t \rightarrow \mathcal{R}_h$ at time $t$. The evolution of this hypergraph over time $T$ captures the dynamic interactions and relationships among entities.
\end{definition}

The hyperedges of HTHGs can dynamically expand to encompass new relationships or contracts to exclude obsolete ones, thereby accurately reflecting the temporal modifications of the system. Then, to harness the analytical power of graph-based algorithms on HTHGs, we propose a novel heterogeneous star expansion strategy that preserves the essential lower-order structures and the directness of connectivity while aptly encapsulating the heterogeneous group interactions. Thus, we offer a refined and effective approach for analyzing complex interaction networks.

Given a $P$-uniform heterogeneous hypergraph $H_P = (V, \mathcal{E}_k)$, the heterogeneous star expansion constructs a heterogeneous graph $G_*=(V_*, E_*, X)$ from $H_P$ by introducing a new node for every heterogeneous hyperedge $e\in \mathcal{E}_k$, thus $V_*=V\cup\mathcal{E}_k$. The new nodes $v\in V_*$ connect each node in the hyperedge $e$, i.e., $E_* = E \cup \{(v, e) \mid v \in e\}$.

By introducing a distinct node for each heterogeneous hyperedge $e \in \mathcal{E}$ and connecting it to nodes within $e$, HTHGN meticulously maintains the heterogeneous nature of the original hypergraph while retaining the connectivity encapsulated in the higher-order relationships of the original hypergraph.

\subsection{Multi-scale Message-passing}

By constructing a $k$-hop/$k$-ring HTHG and conducting the heterogeneous expansion, we obtain the expanded graph $\mathcal{G}_*=\{G_*^t\}_{t=0}^T$. We then utilize a heterogeneous attention message-passing mechanism to aggregate information among heterogeneous nodes as well as between nodes and hyperedges. Specifically, we initialize the features of hyperedge nodes to all zero and execute a single-stage message-passing of the nodes and hyperedges concurrently. This encompasses heterogeneous attention aggregation tailored for heterogeneous relationships within snapshots and temporal attention aggregation designed for dynamics across snapshots.

\paragraph{Heterogeneous Attention Aggregation.}
Heterogeneous attention aggregation is utilized to accomplish message-passing within snapshots of the HTHG. 

For each heterogeneous node $i\in V_*^t$ on snapshot $ G_*^t = (V_*^t, E_*^t, X^t)$, type-preserving attribute projection is  performed through the heterogeneous input layer:
\begin{equation}
    z_i^t = \sigma\left( W_{\phi(i)}x_i^t + b_{\phi(i)}\right),
\end{equation}
where $x_i^t\in\mathbb{R}^{D}$ and $z_i^t\in\mathbb{R}^{d}$ are the original attributes and hidden representation vectors of node $i$; $W_{\phi(i)}\in\mathbb{R}^{d\times D}$ and $b_{\phi(i)}$ are type-dependent learnable transfer matrices and bias vectors; $\sigma(\cdot)$ represents a activation function such as ReLU.

Subsequently, we introduce a relationship type-dependent graph attention mechanism to model distinct semantic relationships within the expanded graph $G_*^t$. Specifically, for neighbor nodes $\mathcal{N}_i^a$ connected to node $i$ under a certain type of relationship $a\in\mathcal{A}_i$, we execute the following $K$-head attention aggregation:
\begin{equation}
\begin{aligned}
    z_{i,a}^t &= \Vert_{k=1}^K\left[\sum_{j\in \mathcal{N}_i^a} \alpha_{ij}^k W_a^k z_i^t \right], \\
    \alpha_{ij}^k &= \text{Softmax}\left(c\cdot\sigma(W_{\phi(i)}^k z_i^t + W_{\phi(j)}^k z_j^t)\right),
\end{aligned}
\end{equation}
where $z_{i,a}^t\in\mathbb{R}^d$ is the attention aggregated representation of all neighbors $j\in\mathcal{N}_i^a$ with respect to relation type $a$; $\alpha_{ij}^k\in\mathbb{R}$ is the normalized mutual attention coefficient of node $i$ with $j$; $W_{a}^k$ and $W_{\phi(i)}^k\in\mathbb{R}^{d\times d}$ are the learnable key and value vectors transfer matrices; $c\in\mathbb{R}^{d}$ is the learnable weight vector for calculating the attention coefficient; $\sigma(\cdot)$ represents a nonlinear activation function such as LeakyReLU.

Following the attention aggregation for neighbor nodes under specific relationship types, we further introduce a self-attention mechanism to aggregate the hidden representations of node $i$ concerning neighbors of different relationship types:
\begin{equation}
\begin{aligned}
\label{eql:input}
    z_{i}^{t} &= \sigma\left(\sum_{a\in\mathcal{A}_i} \beta_a z_{i,a}^t\right), \\
    \beta_a &= \text{Softmax}\left(\frac{1}{|V_*^t|}\sum_{i=1}^{|V_*^t|} q\cdot \text{tanh}(W_a z_{i,a}^t)\right),
\end{aligned}
\end{equation}
where $z_{i}^t$ is the representation of node $i$ under the $t$-th snapshot of expanded HTHG; $\beta_a\in\mathbb{R}$ is the normalized attention score with respect to relationship type $a$; $W_a\in\mathbb{R}^{d\times d}$ and $q\in\mathbb{R}^d$ are learnable attention transformation matrices respectively; $\sigma(\cdot)$ represents the activation function ReLU.

By executing the above heterogeneous attention aggregation within each snapshot, lower-order heterogeneous semantic information is attentively captured, and higher-order interaction and complex semantics are also aggregated into the hyperedge nodes, which ensures a comprehensive integration of both low- and high-order relation, enhancing the overall representation capacity of the hypergraph by encapsulating a broad of interactions and semantics within its structure.

\paragraph{Temporal Attention Aggregation.} This module aggregates node representations calculated under different snapshots and generates dynamic node representations.

Since the temporal attention mechanism will uniformly calculate the node representation under all snapshots, we first add temporal position encoding to each snapshot:
\begin{equation}
    z_{i,p}^t = z_{i}^t + p^t,\quad \\
    p_j^t = \left\{ 
            \begin{aligned}
                &\sin (t/10000^{2j/d}),\quad (\text{if } j \text{ is even}) \\
                &\cos (t/10000^{2j/d}),\quad (\text{if } j \text{ is odd})
            \end{aligned}
            \right.
\end{equation}
where $z_{i,p}^t\in\mathbb{R}^d$ is the hidden representation of node $i$ with position encoding about time $t$; $p^t\in\mathbb{R}^d$ is a deterministic position code with respect to time $t$.

Then, a temporal attention aggregation module is used to aggregate the representations under different snapshots:
\begin{equation}
\begin{aligned}
    \bar{z}_i^t &= \sigma\left(\text{FC}\left( \sum_{t'=1}^{T}(\gamma_i^{t,t'} \cdot W_{\text{V}}z_{i,p}^t))\right)\right), \\
    \gamma_i^{t,t'} &= \text{Softmax}\left(\frac{1}{\sqrt{d}}[W_{\text{K}}z_{i,p}^t]\cdot[W_{\text{Q}}z_{i,p}^{t'}]\right),
\end{aligned}
\end{equation}
where $\bar{z}_i^t\in\mathbb{R}^d$ is the representation vector obtained by attentionally synthesizing each snapshot of node $i$; $\gamma_i^{t,t'}\in\mathbb{R}$ is the normalized mutual attention coefficient between the $t$-th and $t'$-th snapshots of node $i$; $W_{\text{K}} \in\mathbb{R}^{d\times d}$, $W_{\text{Q}} \in\mathbb{R}^{d\times d}$ and $W_{\text{V}} \in\mathbb{R}^{d\times d}$ are learnable transformation matrices of Key, Query and Value vectors used to calculate attention weights; FC$(\cdot)$ is a trainable fully connected layer. The above parameters are not shared among different target node types $\phi(i)$.

Then, a heterogeneous gated residual connection mechanism is used to connect with the previous layer input and calculate the final node representation by summing all snapshots.
\begin{equation}
    \hat{z}_i = \sum_{t=1}^T \left( r_{\phi(i)}^t\cdot\bar{z}_i^t + (1-r_{\phi(i)}^t)\text{ FC}(z_i^{t}) \right), \\
\end{equation}
where $z_i^{t}$ and $\hat{z}_i$ are the residual node attribute vector obtained by Equation~(\ref{eql:input}) and the updated node representation vector updated by the whole message-passing mechanism; $r_{\phi(i)}\in\mathbb{R}$ a is a trainable variable used to control the update strength of node $i$; FC$(\cdot)$ is a trainable fully connected layer.

Stacking two or more layers of the above attention modules enables a single-stage message-passing process: simultaneously from heterogeneous nodes to hyperedge nodes and back to the heterogeneous nodes. This layered approach enhances the depth of information integration, allowing for the iterative refinement of node representations and facilitates a comprehensive bidirectional flow of information. This bidirectional message passing harnesses the strengths of both direct and higher-order interactions, enhancing the ability to capture and understand the complex, multi-faceted relationships present within the hypergraph structure.

\subsection{Heterogeneous Contrastive Optimization}

To enable the HTHGN to learn heterogeneous semantic information from network data adaptively and to circumvent the issue of lower-order structural information loss caused by the introduction of hyperedges, we optimize the entire model through a self-supervised contrastive learning objective. For the given HTHG $\mathcal{G}_* = \{G_*^t\}_{t=1}^T$ and the target node $i\in V$, we select its heterogeneous neighbors at following snapshot as positive sample set $\mathbb{P}_i^{T+1}$ and uniformly sample $Q$ non-neighbors as negative sample set $\mathbb{N}_i^{T+1}$:
\begin{equation}
\begin{aligned}
    \mathbb{P}_i^{T+1} = \{u \mid u \in V^{T+1} \land u \in \mathcal{N}_i^{T+1}\},\quad\\
    \mathbb{N}_i^{T+1} = \{v \mid v \in V^{T+1} \land v \notin \mathcal{N}_i^{T+1} \land v\neq i\}, \\
\end{aligned}
\end{equation}
Subsequently, we employ a projection head as a discriminator to assess the likelihood of the existence of lower-order relationships between node pairs in the $T+1$-th snapshot:
\begin{equation}
    \mathcal{D}(\hat{z}_i, \hat{z}_j) = \text{FC }(\sigma(\text{FC }(\hat{z}_i \Vert \hat{z}_j))).
\end{equation}

We draw inspiration from the Deep InfoMax for the objective function, adopting a noise-contrastive estimation framework paired with a binary cross-entropy (BCE) loss. This loss function discriminates between pairs of samples originating from the joint distribution of nodes and their corresponding heterogeneous neighbors (positive examples) and those from the marginal distributions (negative examples):
\begin{equation}
\begin{aligned}
    \mathcal{L} =      \sum_{i\in V^t} \bigg(     &\sum_{j\in\mathbb{P}_i^{T+1}}      \mathbb{E}\left[\log \mathcal{D}(\hat{z}_i, \hat{z}_j)\right] \\     &+ \sum_{j\in\mathbb{N}_i^{T+1}}      \mathbb{E} \left[          \log\left( 1 - \mathcal{D}(\hat{z}_i, \hat{z}_j)\right)     \right] \bigg).
\end{aligned}
\end{equation}

The model is effectively trained to distinguish between authentic neighboring node relationships and unrelated node pairs across snapshots, thereby enhancing its ability to infer and preserve lower-order connections.

\section{Experiments}
\label{sec:experiments}

\begin{table*}[!htb]
\resizebox{\textwidth}{!}{%
\begin{tabular}{cccccccc}
\hline
Dataset      & \multicolumn{2}{c}{Yelp}            & \multicolumn{2}{c}{DBLP}            & \multicolumn{2}{c}{AMiner}                & \multirow{2}{*}{Avg. Rank} \\ \cline{1-7}
Metrics      & AUC              & AP               & AUC              & AP               & AUC              & AP                     &                            \\ \hline
VGAE         & 58.62 $\pm$ 4.79 & 59.71 $\pm$ 4.48 & 77.40 $\pm$ 1.41 & 80.55 $\pm$ 1.53 & 84.56 $\pm$ 2.68 & 87.13 $\pm$ 2.53       & 9.67                       \\
GATv2        & 59.87 $\pm$ 1.31 & 57.20 $\pm$ 1.76 & 83.28 $\pm$ 0.13 & 84.71 $\pm$ 0.27 & 89.12 $\pm$ 0.30 & 90.60 $\pm$ 0.40       & 6.17                       \\
DGI          & 55.68 $\pm$ 3.11 & 57.36 $\pm$ 2.66 & 73.36 $\pm$ 2.93 & 77.65 $\pm$ 2.47 & 80.71 $\pm$ 5.59 & 84.04 $\pm$ 4.60       & 12.33                      \\
EvolveGCN    & 54.85 $\pm$ 5.51 & 54.79 $\pm$ 4.07 & 71.26 $\pm$ 6.87 & 75.33 $\pm$ 5.70 & 74.90 $\pm$ 7.87 & 78.97 $\pm$ 6.28       & 14.67                      \\
DySAT        & 61.88 $\pm$ 2.68 & 58.57 $\pm$ 2.71 & 78.61 $\pm$ 1.54 & 80.56 $\pm$ 1.42 & 83.76 $\pm$ 0.98 & 85.31 $\pm$ 1.18       & 9.50                       \\
HyperGCN     & 59.18 $\pm$ 2.05 & 55.64 $\pm$ 2.03 & 72.60 $\pm$ 1.04 & 73.96 $\pm$ 0.68 & 75.07 $\pm$ 1.24 & 75.64 $\pm$ 2.31       & 13.67                      \\
UniGCN       & 57.47 $\pm$ 5.37 & 54.99 $\pm$ 4.13 & 75.14 $\pm$ 0.76 & 74.45 $\pm$ 3.26 & 82.35 $\pm$ 0.62 & 81.50 $\pm$ 1.45       & 12.67                      \\
UniGAT       & 55.47 $\pm$ 0.57 & 52.01 $\pm$ 0.69 & 83.88 $\pm$ 0.31 & 86.54 $\pm$ 0.34 & 89.13 $\pm$ 0.51 & {\ul 91.19 $\pm$ 0.62} & 7.00                       \\
HGNNP        & 62.16 $\pm$ 3.80 & 60.14 $\pm$ 3.54 & 80.39 $\pm$ 0.20 & 83.39 $\pm$ 0.38 & 85.59 $\pm$ 0.82 & 88.19 $\pm$ 0.71       & 7.17                       \\ \hline
metapath2vec & 63.79 $\pm$ 0.41 & 59.28 $\pm$ 0.46 & 64.28 $\pm$ 2.15 & 60.28 $\pm$ 2.56 & 71.05 $\pm$ 1.61 & 68.95 $\pm$ 1.87       & 13.00                      \\
R-GCN        & 52.72 $\pm$ 2.25 & 51.66 $\pm$ 1.48 & 82.28 $\pm$ 3.79 & 84.18 $\pm$ 3.95 & 88.16 $\pm$ 1.55 & 89.66 $\pm$ 1.47       & 9.83                       \\
HGT          & 55.86 $\pm$ 1.97 & 54.07 $\pm$ 2.00 & 82.32 $\pm$ 0.46 & 85.32 $\pm$ 0.55 & 87.27 $\pm$ 0.63 & 89.94 $\pm$ 0.53       & 8.33                       \\
HetSANN-GVAE & 59.28 $\pm$ 2.41 & 57.43 $\pm$ 2.34 & 81.51 $\pm$ 1.53 & 85.08 $\pm$ 2.44 & 87.52 $\pm$ 0.55 & 90.19 $\pm$ 0.80       & 6.83                       \\
HPN          & 62.02 $\pm$ 1.04 & 60.24 $\pm$ 1.53 & 81.30 $\pm$ 1.12 & 82.64 $\pm$ 1.37 & 84.39 $\pm$ 1.06 & 87.97 $\pm$ 0.77       & 7.67                       \\
DyHATR       & 63.58 $\pm$ 1.37 & 63.60 $\pm$ 1.29 & 69.61 $\pm$ 1.64 & 69.82 $\pm$ 1.72 & 75.90 $\pm$ 2.51 & 76.80 $\pm$ 2.43       & 11.33                      \\
HTGNN &
  {\ul 70.43 $\pm$ 3.36} &
  {\ul 67.45 $\pm$ 4.18} &
  {\ul 85.94 $\pm$ 3.47} &
  {\ul 87.17 $\pm$ 3.30} &
  {\ul 90.50 $\pm$ 3.33} &
  90.81 $\pm$ 3.42 &
  {\ul 2.17} \\ \hline
HTHGN &
  \textbf{74.04 $\pm$ 4.82} &
  \textbf{89.56 $\pm$ 3.01} &
  \textbf{91.33 $\pm$ 1.61} &
  \textbf{96.97 $\pm$ 0.56} &
  \textbf{96.58 $\pm$ 1.14} &
  \textbf{98.80 $\pm$ 0.40} &
  \textbf{1.00} \\ \hline
\end{tabular}%
}
\caption{AUC and AP scores of link prediction tasks between HTHGN and baselines in three datasets.}
\label{tab:linkpred}
\end{table*}

\subsection{Datasets and Baselines}
This section evaluates the proposed HTHGN and baselines on three real-world datasets: Yelp, DBLP, and AMiner. We compare static homogeneous methods \textbf{VGAE}~\cite{kipf2016variational}, \textbf{GATv2}~\cite{brody2022how}, \textbf{DGI}~\cite{veličković2018deep}; dynamic homogeneous GNNs \textbf{EvolveGCN}~\cite{EvolveGCN2020}, \textbf{DySAT}~\cite{SankarDySAT}; hypergraph methods \textbf{HyperGCN}~\cite{10.5555/3454287.3454422}, \textbf{UniGCN} and \textbf{UniGAT}~\cite{ijcai2021p353} , \textbf{HGNNP}~\cite{9795251}; static heterogeneous methods \textbf{metapath2vec}~\cite{10.1145/3097983.3098036}, \textbf{R-GCN}~\cite{SchlichtkrullRGCN}, \textbf{HGT}~\cite{10.1145/3366423.3380027}, \textbf{H-GVAE}~\cite{dalvi2022variational}, \textbf{HPN}~\cite{9428609}; dynamic heterogeneous GNNs \textbf{DyHATR}~\cite{10.1007/978-3-030-67658-2_17} and \textbf{HTGNN}~\cite{doi:10.1137/1.9781611977172.74}. 

\subsection{Experiment Setup}
We conducted dynamic link prediction and new link prediction experiments to verify the gain of higher-order interactions on representation learning performance. We held out the last $3$ snapshots for testing and trained the model on the remaining snapshots. The link prediction uses all edges in $T+1$-th snapshot as positive edges, while the new link prediction only evaluates edges that have not appeared. We performed $5$ repeated randomized experiments for all methods and reported their means and standard deviations. More setup and implementation details see Appendix~B.

\subsection{Experiment Results}
\label{sec:mainresults}

\begin{figure*}[!htb]
    \centering
    \begin{minipage}[t]{0.48\textwidth}
        \begin{subfigure}[b]{0.49\linewidth}
            \includegraphics[width=\linewidth]{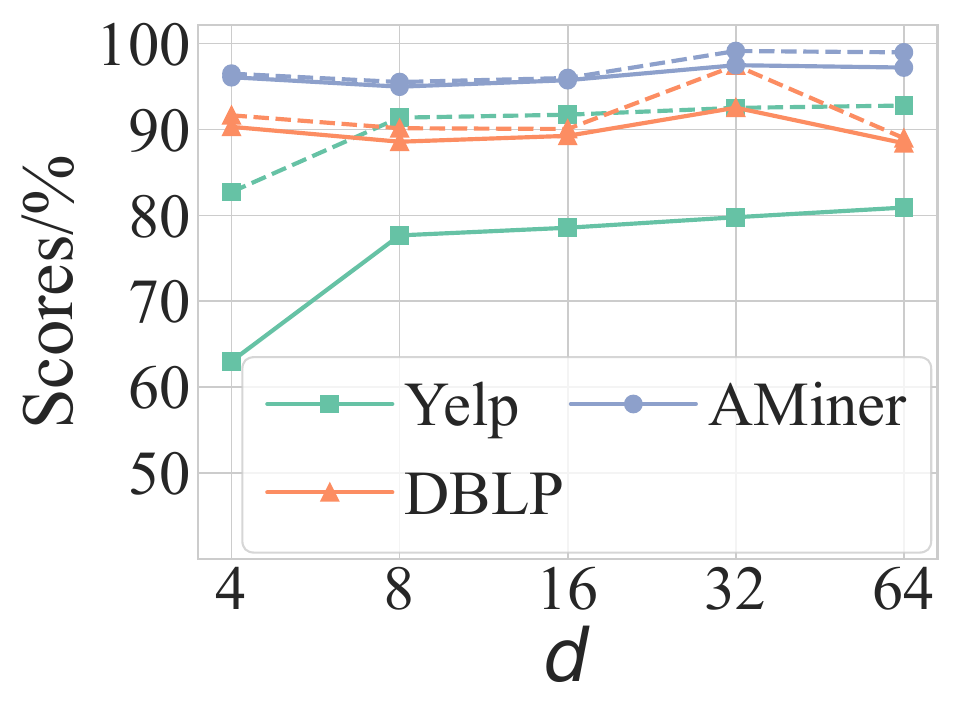}
            \caption{Link Prediction}
        \end{subfigure}
        \hfill
        \begin{subfigure}[b]{0.49\linewidth}
            \includegraphics[width=\linewidth]{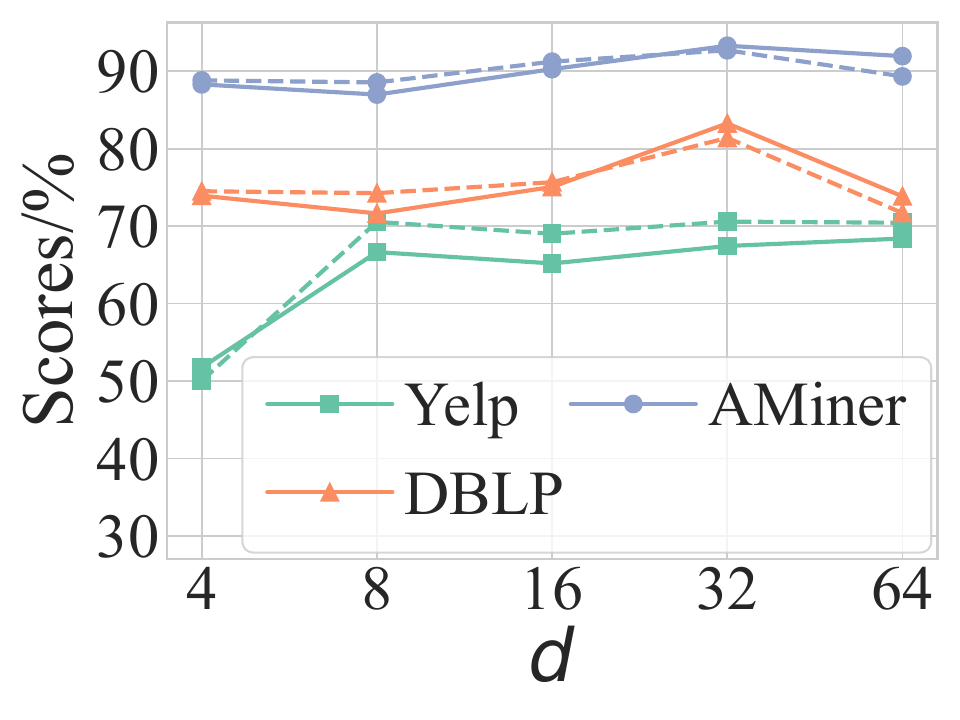}
            \caption{New Link Prediction}
        \end{subfigure}
    \caption{Impact of dimension.}
    \label{fig:dim}
    \end{minipage}
    \hfill
    \begin{minipage}[t]{0.48\textwidth}
        \begin{subfigure}[b]{0.49\linewidth}
            \includegraphics[width=\linewidth]{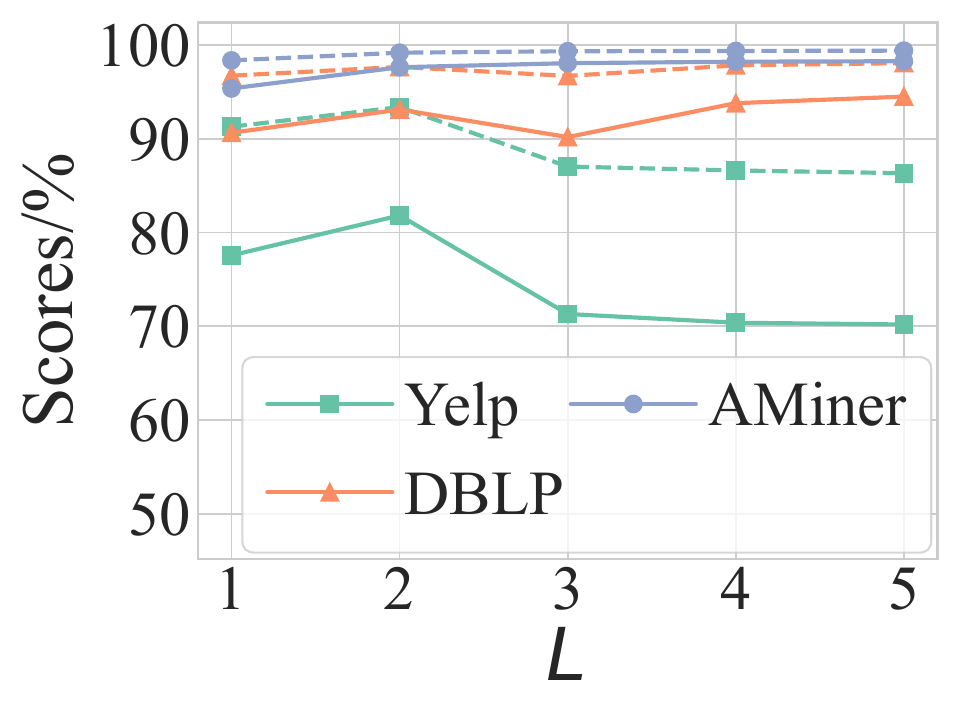}
            \caption{Link Prediction}
        \end{subfigure}
        \hfill
        \begin{subfigure}[b]{0.49\linewidth}
            \includegraphics[width=\linewidth]{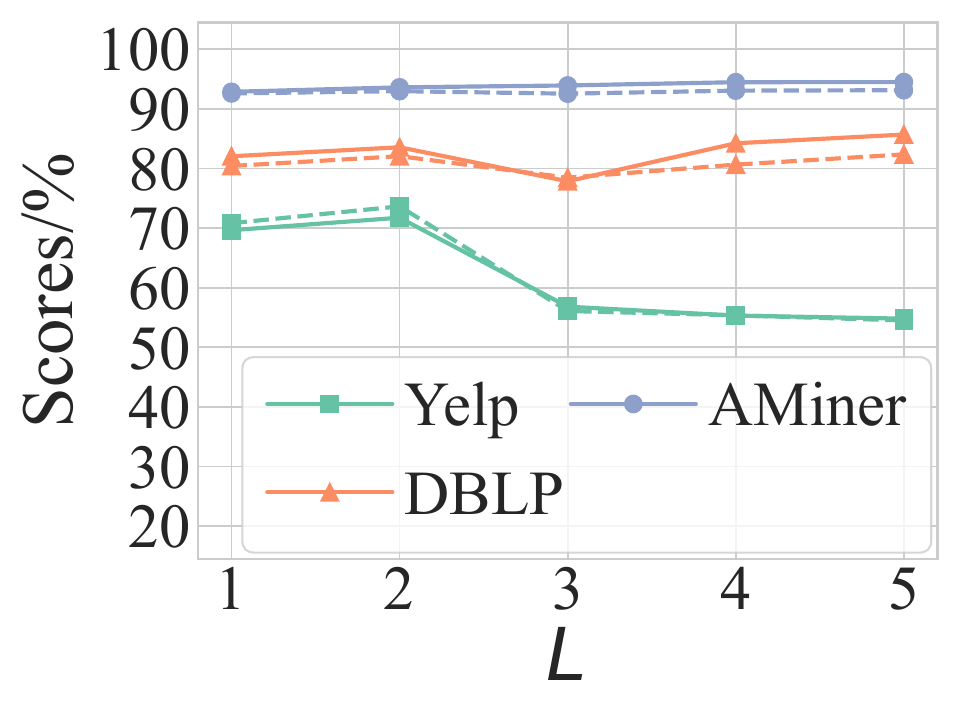}
            \caption{New Link Prediction}
        \end{subfigure}
    \caption{Impact of layer.}
    \label{fig:layer}
    \end{minipage}
\end{figure*}

\paragraph{Link Prediction}
Our comparative experimental results are summarized in Table~\ref{tab:linkpred}, and more results are in Appendix~E. The results show that HTHGN achieves excellent performance on both AUC and AP metrics in all datasets. In particular, we note that methods designed for heterogeneous graphs generally outperform homogeneous graph methods, demonstrating the clear gains of introducing higher-order heterogeneous type information into representation learning. We believe this is because more semantic relatedness can be captured through type information, where homogeneous graph methods are limited. Furthermore, it should be noted that although models designed for homogeneous hypergraphs, such as UniGAT and HGNNP, enlarge the receptive field, their performance suffers due to more noise and the inability to model semantic information. We further observe that compared to static heterogeneous graph methods, dynamic heterogeneous graph methods achieve more competitive performance, thus validating the advantages of modeling network temporal evolution. Compared with the novel HTGNN, our method achieves better results by modeling heterogeneous high-order interaction. We believe this can be attributed to the sparsity of the network structure, and HTHGN rescues this through heterogeneous hyperedges and significantly improves the attention encoder performance. Similar and more significant experimental conclusions can be drawn from the more challenging new link prediction experiment, which is reported in Appendix~E.1.

\begin{figure*}[!htb]
    \centering
    \begin{minipage}[t]{0.48\textwidth}
    \begin{subfigure}[b]{0.49\linewidth}
        \includegraphics[width=\linewidth]{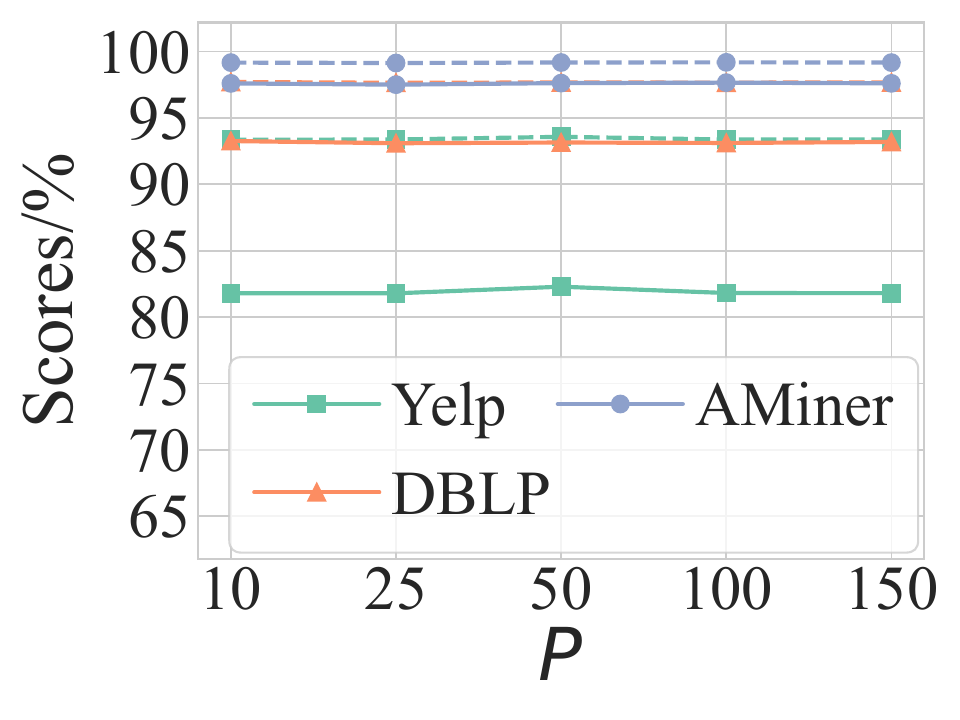}
        \caption{Link Prediction}
    \end{subfigure}
    \hfill
    \begin{subfigure}[b]{0.49\linewidth}
        \includegraphics[width=\linewidth]{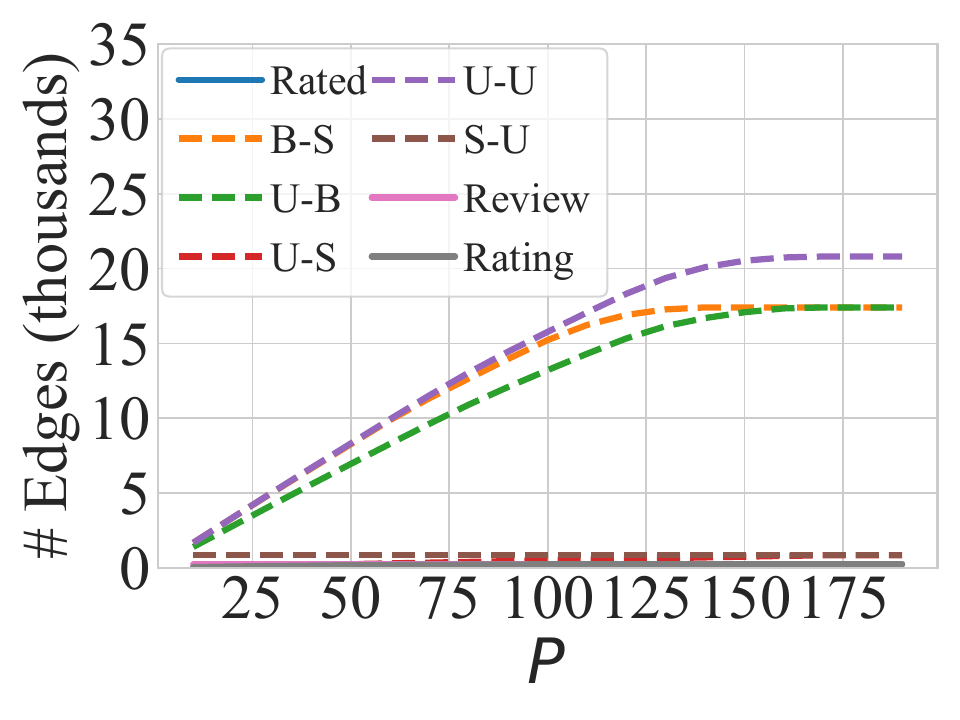}
        \caption{Yelp Dataset}
    \end{subfigure}
    \caption{Impact of $P$-uniform on HTHGN. (a) shows AUC and AP scores; (b) shows the number of low- and high-order hyperedges on the Yelp dataset. Results indicate that increasing $P$ enlarges the hypergraph but HTHGN maintains stable performance.}
    \label{fig:uniform}
    \end{minipage}
    \hfill
    \begin{minipage}[t]{0.49\textwidth}
        \includegraphics[width=\linewidth]{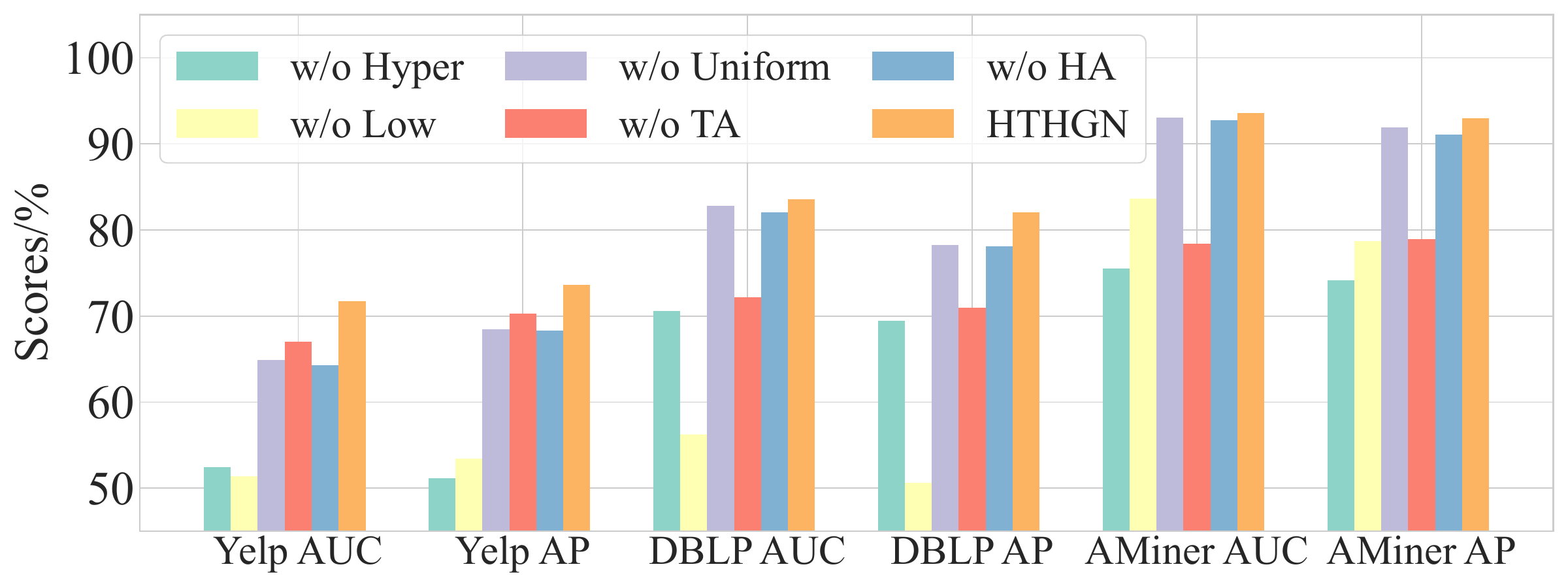}
        \caption{Ablation results of HTHGN and its five ablated variants on three datasets (Yelp, DBLP, and AMiner) with AUC and AP as evaluation metrics. The results demonstrate the performance contribution of each component in the model.}
        \label{fig:ablation-nlp}
    \end{minipage}
\end{figure*}

\paragraph{Impact of Hypergraph Construction} 
For the $k$-hop/$k$-ring hyperedge construction method proposed in this paper, we set different values for $k$ and compare its performance on 3 datasets under the HTHGN model. From Table~4 in Appendix~E.2, we can observe that when $k=1$, the performance gap between $1$-hop and $1$-ring is not obvious. This may be because when $k=1$, the hyperedge is equivalent to its direct neighbors, so there is no functional difference between $1$-hop and $1$-ring hyperedges. In addition, we should also notice that in all 3 datasets, the model performance gradually increases when $k$ increases. This reflects that increasing the group interaction receptive field for HTHGN can enrich the heterogeneous relationship semantic and thereby improve performance. It should be noted that in all 3 datasets, the HTHGN of $3$-ring hyperedge is significantly better than that of $3$-hop hyperedge. We believe that due to the $2$-layer design of the HTHGN encoder, $2$-order neighbors can complete message-passing through low-order interactions. Therefore, the $3$-ring hyperedge focuses on heterogeneous nodes with a distance of exactly $3$, which can bring more pure and inspiring high-order interactive information.

\paragraph{Impact of $P$-uniform} 
To explore the impact of P-uniform on HTHGN representation learning performance, we analyzed the hyperedge numbers and model performance under different values of $P$, which are reported in Figure~\ref{fig:uniform}. It can be seen from the results that as $P$ increases, the size of the hyperedge in HTHG increases sharply. This is because the expansion of the hypergraph often leads to a significant increase in the number of nodes and edges. On the contrary, $P$-uniform HTHG usually has fewer nodes and edges in the converted ordinary graph, which can balance the contradiction between model performance and computing resources and improve computing efficiency. In addition, it is worth noting that the model performance of the proposed HTHGN is relatively stable under different values of $P$. This result shows the proposed HTHGN can also perform constantly well under a smaller hyperedge scale, verifying that high-order relationships are important for improving the general validity of learning performance. 
Besides, this result demonstrates that compared with the naive cluster/star expansion algorithm, HTHGN can significantly improve computational efficiency by controlling the $P$-uniform hyperedges.

\paragraph{Parameter Sensitive Analysis} 
Here, we analyze the impact of the main hyperparameters in HTHGN on model performance. As shown in Figure~\ref{fig:dim} and~\ref{fig:layer}, HTHGN's performance fluctuates slightly under different configurations, which shows that the model is stable on most tasks and datasets.

\paragraph{Ablation Study}
To verify the effectiveness of each module, we performed ablation experiments and reported the results as shown in Figure~\ref{fig:ablation-nlp} and Appendix~E.4. Among them, \textit{w/o Hyper} means removing the hypergraph structure, and \textit{w/o Low} means removing the low-order structure. Both of them significantly affect the performance of each dataset. \textit{w/o Uniform} uses ununiformed hyperedges. \textit{w/o TA} and \textit{w/o HA} respectively represent the removal of the Temporal and Heterogeneous Attention Aggregation modules. Their impact on performance verifies that both temporal dependence and heterogeneous semantics are indispensable.

\section{Related Works}
\label{sec:related}

\textbf{Graph representation learning.} Real-world complex networks containing heterogeneity and dynamics are ubiquitous, and representation learning and link prediction about them are usually divided into two separate research directions. On the one hand, methods for heterogeneity modeling include meta-path-based methods~\cite{10.1145/3308558.3313562,9428609,Yang_Yan_Pan_Ye_Fan_2023,10.1145/3485189,9091208} and heterogeneous message-passing-based methods~\cite{SchlichtkrullRGCN,10.1145/3366423.3380027,dalvi2022variational,10.1145/3543507.3583493,doi:10.1137/1.9781611977172.74}. On the other hand, methods for dynamics include decomposition-based methods~\cite{ijcai2017p467,MA2017361}, temporal random walk-based methods~\cite{Liu_Zhou_Zhu_Gu_He_2020,10.1145/3552326.3567491}, and deep learning-based methods~\cite{EvolveGCN2020,10436338}. However, these methods cannot simultaneously effectively capture the dynamics, heterogeneity, and high-order interaction. 

\textbf{Hypergraph representation learning.} Hypergraph representation learning aims to embed the hypergraph into a low-dimensional space and maintain the original hypergraph structural information~\cite{10.1145/3605776}. Early hypergraph representation learning methods were based on spectral theory~\cite{NIPS2006_dff8e9c2,10.5555/3504035.3504523} and structure-preserving~\cite{9169850,mlg2020_15} learning node representations. However, these methods are shallow and cannot model highly nonlinear relationships. The advent of GNN-based methods offered a breakthrough with their end-to-end learning and scalability~\cite{10.1145/3605776,10.5555/3454287.3454422,ijcai2021p353,9795251,Jiao_Chen_Bao_Zhang_Wu_2024}, which define the hypergraph Laplacian and train classic GNNs on hypergraphs. Besides, GNNs designed for heterogeneous hypergraphs usually use hyperedge type~\cite{8594913,10.1145/3437963.3441835,LU2023109818} or meta-path~\cite{9415142,doi:10.1137/1.9781611977653.ch80} to decompose the heterogeneous hypergraph into different semantic relations. This dependency on specific prior knowledge and the static nature limits their efficiency in capturing the dynamic within hypergraphs.

\section{Conclusions}
\label{sec:con}
In this paper, we propose a HTHGN method to construct and learn heterogeneous high-order interactions in dynamic heterogeneous graphs without additional knowledge. To better divide different receptive fields, we define and analyze two different types of heterogeneous uniform hyperedge construction methods. The effectiveness of the HTHGN proposed in this paper is verified through extensive experiments on three real-world datasets. The limitation of this work is that the process of HTHGN modeling high-order semantic information is relatively complex, and the interpretability of its effectiveness still needs to be explored.

\section*{Acknowledgments}
This work was supported in part by the Zhejiang Provincial Natural Science Foundation of China under Grants LDT23F01015F01 and LMS25F030011, in part by the Key Technology Research and Development Program of Zhejiang Province under Grant 2025C1023, and in part by the National Natural Science Foundation of China under Grant 62372146.

\bibliographystyle{named}
\bibliography{ijcai25}

\begin{thebibliography}{}

\bibitem[\protect\citeauthoryear{Antelmi \bgroup \em et al.\egroup }{2023}]{10.1145/3605776}
Alessia Antelmi, Gennaro Cordasco, Mirko Polato, Vittorio Scarano, Carmine Spagnuolo, and Dingqi Yang.
\newblock A survey on hypergraph representation learning.
\newblock {\em ACM Computing Surveys}, 56(1), 2023.

\bibitem[\protect\citeauthoryear{Barros \bgroup \em et al.\egroup }{2021}]{10.1145/3483595}
Claudio D.~T. Barros, Matheus R.~F. Mendon\c{c}a, Alex~B. Vieira, and Artur Ziviani.
\newblock A survey on embedding dynamic graphs.
\newblock {\em ACM Computing Surveys}, 55(1), 2021.

\bibitem[\protect\citeauthoryear{Baytas \bgroup \em et al.\egroup }{2018}]{8594913}
Inci~M. Baytas, Cao Xiao, Fei Wang, Anil~K. Jain, and Jiayu Zhou.
\newblock Heterogeneous hyper-network embedding.
\newblock In {\em 2018 IEEE International Conference on Data Mining (ICDM)}, pages 875--880, 2018.

\bibitem[\protect\citeauthoryear{Brody \bgroup \em et al.\egroup }{2022}]{brody2022how}
Shaked Brody, Uri Alon, and Eran Yahav.
\newblock How attentive are graph attention networks?
\newblock In {\em ICLR}, 2022.

\bibitem[\protect\citeauthoryear{Dalvi \bgroup \em et al.\egroup }{2022}]{dalvi2022variational}
Abhishek Dalvi, Ayan Acharya, Jing Gao, and Vasant~G Honavar.
\newblock Variational graph auto-encoders for heterogeneous information network.
\newblock In {\em NeurIPS 2022 Workshop: New Frontiers in Graph Learning}, 2022.

\bibitem[\protect\citeauthoryear{Dong \bgroup \em et al.\egroup }{2017}]{10.1145/3097983.3098036}
Yuxiao Dong, Nitesh~V. Chawla, and Ananthram Swami.
\newblock Metapath2vec: Scalable representation learning for heterogeneous networks.
\newblock In {\em Proceedings of the 23rd ACM SIGKDD International Conference on Knowledge Discovery and Data Mining}, page 135–144, 2017.

\bibitem[\protect\citeauthoryear{Fan \bgroup \em et al.\egroup }{2022}]{doi:10.1137/1.9781611977172.74}
Yujie Fan, Mingxuan Ju, Chuxu Zhang, and Yanfang Ye.
\newblock Heterogeneous temporal graph neural network.
\newblock In {\em Proceedings of the 2022 SIAM International Conference on Data Mining}, pages 657--665, 2022.

\bibitem[\protect\citeauthoryear{Fang \bgroup \em et al.\egroup }{2022}]{10.1145/3485189}
Yang Fang, Xiang Zhao, Peixin Huang, Weidong Xiao, and Maarten de~Rijke.
\newblock Scalable representation learning for dynamic heterogeneous information networks via metagraphs.
\newblock {\em ACM Transactions on Information Systems}, 40(4), 2022.

\bibitem[\protect\citeauthoryear{Gao \bgroup \em et al.\egroup }{2023a}]{9795251}
Yue Gao, Yifan Feng, Shuyi Ji, and Rongrong Ji.
\newblock Hgnn+: General hypergraph neural networks.
\newblock {\em IEEE Transactions on Pattern Analysis and Machine Intelligence}, 45(3):3181--3199, 2023.

\bibitem[\protect\citeauthoryear{Gao \bgroup \em et al.\egroup }{2023b}]{gao2023hierarchical}
Ziqi Gao, Chenran Jiang, Jiawen Zhang, Xiaosen Jiang, Lanqing Li, Peilin Zhao, Huanming Yang, Yong Huang, and Jia Li.
\newblock Hierarchical graph learning for protein--protein interaction.
\newblock {\em Nature Communications}, 14(1):1093, 2023.

\bibitem[\protect\citeauthoryear{Hamilton \bgroup \em et al.\egroup }{2017}]{10.5555/3294771.3294869}
William~L. Hamilton, Rex Ying, and Jure Leskovec.
\newblock Inductive representation learning on large graphs.
\newblock In {\em NeurIPS}, page 1025–1035, 2017.

\bibitem[\protect\citeauthoryear{Hu \bgroup \em et al.\egroup }{2020}]{10.1145/3366423.3380027}
Ziniu Hu, Yuxiao Dong, Kuansan Wang, and Yizhou Sun.
\newblock Heterogeneous graph transformer.
\newblock In {\em Proceedings of The Web Conference 2020}, page 2704–2710, 2020.

\bibitem[\protect\citeauthoryear{Huan \bgroup \em et al.\egroup }{2023}]{10.1145/3552326.3567491}
Chengying Huan, Shuaiwen~Leon Song, Santosh Pandey, Hang Liu, Yongchao Liu, Baptiste Lepers, Changhua He, Kang Chen, Jinlei Jiang, and Yongwei Wu.
\newblock Tea: A general-purpose temporal graph random walk engine.
\newblock In {\em Proceedings of the Eighteenth European Conference on Computer Systems}, page 182–198, 2023.

\bibitem[\protect\citeauthoryear{Huang and Yang}{2021}]{ijcai2021p353}
Jing Huang and Jie Yang.
\newblock Unignn: a unified framework for graph and hypergraph neural networks.
\newblock In {\em IJCAI}, pages 2563--2569, 2021.

\bibitem[\protect\citeauthoryear{Ji \bgroup \em et al.\egroup }{2023}]{9428609}
Houye Ji, Xiao Wang, Chuan Shi, Bai Wang, and Philip~S. Yu.
\newblock Heterogeneous graph propagation network.
\newblock {\em IEEE Transactions on Knowledge and Data Engineering}, 35(1):521--532, 2023.

\bibitem[\protect\citeauthoryear{Jiao \bgroup \em et al.\egroup }{2024}]{Jiao_Chen_Bao_Zhang_Wu_2024}
Pengfei Jiao, Hongqian Chen, Qing Bao, Wang Zhang, and Huaming Wu.
\newblock Enhancing multi-scale diffusion prediction via sequential hypergraphs and adversarial learning.
\newblock {\em Proceedings of the AAAI Conference on Artificial Intelligence}, 38(8):8571--8581, 2024.

\bibitem[\protect\citeauthoryear{Kipf and Welling}{2016}]{kipf2016variational}
Thomas~N Kipf and Max Welling.
\newblock Variational graph auto-encoders.
\newblock {\em arXiv preprint arXiv:1611.07308}, 2016.

\bibitem[\protect\citeauthoryear{Kipf and Welling}{2017}]{kipf2017semisupervised}
Thomas~N. Kipf and Max Welling.
\newblock Semi-supervised classification with graph convolutional networks.
\newblock In {\em ICLR}, 2017.

\bibitem[\protect\citeauthoryear{Li \bgroup \em et al.\egroup }{2023}]{9415142}
Jianxin Li, Hao Peng, Yuwei Cao, Yingtong Dou, Hekai Zhang, Philip~S. Yu, and Lifang He.
\newblock Higher-order attribute-enhancing heterogeneous graph neural networks.
\newblock {\em IEEE Transactions on Knowledge \& Data Engineering}, 35(01):560--574, 2023.

\bibitem[\protect\citeauthoryear{Liu \bgroup \em et al.\egroup }{2020}]{Liu_Zhou_Zhu_Gu_He_2020}
Zhining Liu, Dawei Zhou, Yada Zhu, Jinjie Gu, and Jingrui He.
\newblock Towards fine-grained temporal network representation via time-reinforced random walk.
\newblock {\em Proceedings of the AAAI Conference on Artificial Intelligence}, 34(04):4973--4980, 2020.

\bibitem[\protect\citeauthoryear{Liu \bgroup \em et al.\egroup }{2024}]{10436338}
Huan Liu, Pengfei Jiao, Xuan Guo, Huaming Wu, Mengzhou Gao, and Jilin Zhang.
\newblock Hgn2t: A simple but plug-and-play framework extending hgnns on heterogeneous temporal graphs.
\newblock {\em IEEE Transactions on Big Data}, 10(5):620--632, 2024.

\bibitem[\protect\citeauthoryear{Lu \bgroup \em et al.\egroup }{2023}]{LU2023109818}
Yifan Lu, Mengzhou Gao, Huan Liu, Zehao Liu, Wei Yu, Xiaoming Li, and Pengfei Jiao.
\newblock Neighborhood overlap-aware heterogeneous hypergraph neural network for link prediction.
\newblock {\em Pattern Recognition}, 144:109818, 2023.

\bibitem[\protect\citeauthoryear{Ma \bgroup \em et al.\egroup }{2017}]{MA2017361}
Xiaoke Ma, Penggang Sun, and Guimin Qin.
\newblock Nonnegative matrix factorization algorithms for link prediction in temporal networks using graph communicability.
\newblock {\em Pattern Recognition}, 71:361--374, 2017.

\bibitem[\protect\citeauthoryear{Mao \bgroup \em et al.\egroup }{2023}]{10.1145/3543507.3583493}
Qiheng Mao, Zemin Liu, Chenghao Liu, and Jianling Sun.
\newblock Hinormer: Representation learning on heterogeneous information networks with graph transformer.
\newblock In {\em Proceedings of the ACM Web Conference}, page 599–610, 2023.

\bibitem[\protect\citeauthoryear{Pareja \bgroup \em et al.\egroup }{2020}]{EvolveGCN2020}
Aldo Pareja, Giacomo Domeniconi, Jie Chen, Tengfei Ma, Toyotaro Suzumura, Hiroki Kanezashi, Tim Kaler, Tao Schardl, and Charles Leiserson.
\newblock Evolvegcn: Evolving graph convolutional networks for dynamic graphs.
\newblock In {\em AAAI}, volume~34, pages 5363--5370, 2020.

\bibitem[\protect\citeauthoryear{Saito \bgroup \em et al.\egroup }{2018}]{10.5555/3504035.3504523}
Shota Saito, Danilo~P Mandic, and Hideyuki Suzuki.
\newblock Hypergraph p-laplacian: a differential geometry view.
\newblock In {\em AAAI}, 2018.

\bibitem[\protect\citeauthoryear{Sankar \bgroup \em et al.\egroup }{2020}]{SankarDySAT}
Aravind Sankar, Yanhong Wu, Liang Gou, Wei Zhang, and Hao Yang.
\newblock Dysat: Deep neural representation learning on dynamic graphs via self-attention networks.
\newblock In {\em Proceedings of the 13th International Conference on Web Search and Data Mining}, page 519–527, 2020.

\bibitem[\protect\citeauthoryear{Schlichtkrull \bgroup \em et al.\egroup }{2018}]{SchlichtkrullRGCN}
Michael Schlichtkrull, Thomas~N. Kipf, Peter Bloem, Rianne van den Berg, Ivan Titov, and Max Welling.
\newblock Modeling relational data with graph convolutional networks.
\newblock In {\em The Semantic Web}, pages 593--607, 2018.

\bibitem[\protect\citeauthoryear{Sun \bgroup \em et al.\egroup }{2021}]{10.1145/3437963.3441835}
Xiangguo Sun, Hongzhi Yin, Bo~Liu, Hongxu Chen, Jiuxin Cao, Yingxia Shao, and Nguyen~Quoc Viet~Hung.
\newblock Heterogeneous hypergraph embedding for graph classification.
\newblock In {\em Proceedings of the 14th ACM International Conference on Web Search and Data Mining}, page 725–733, 2021.

\bibitem[\protect\citeauthoryear{Sybrandt and Safro}{2020}]{mlg2020_15}
Justin Sybrandt and Ilya Safro.
\newblock First and higher-order bipartite embeddings.
\newblock In {\em Proceedings of the 16th International Workshop on Mining and Learning with Graphs (MLG)}, 2020.

\bibitem[\protect\citeauthoryear{Sybrandt \bgroup \em et al.\egroup }{2022}]{9169850}
Justin Sybrandt, Ruslan Shaydulin, and Ilya Safro.
\newblock Hypergraph partitioning with embeddings.
\newblock {\em IEEE Transactions on Knowledge and Data Engineering}, 34(6):2771--2782, 2022.

\bibitem[\protect\citeauthoryear{Veličković \bgroup \em et al.\egroup }{2019}]{veličković2018deep}
Petar Veličković, William Fedus, William~L. Hamilton, Pietro Liò, Yoshua Bengio, and R~Devon Hjelm.
\newblock Deep graph infomax.
\newblock In {\em ICLR}, 2019.

\bibitem[\protect\citeauthoryear{Wang \bgroup \em et al.\egroup }{2019}]{10.1145/3308558.3313562}
Xiao Wang, Houye Ji, Chuan Shi, Bai Wang, Yanfang Ye, Peng Cui, and Philip~S Yu.
\newblock Heterogeneous graph attention network.
\newblock In {\em The World Wide Web Conference}, page 2022–2032, 2019.

\bibitem[\protect\citeauthoryear{Wang \bgroup \em et al.\egroup }{2022}]{9091208}
Xiao Wang, Yuanfu Lu, Chuan Shi, Ruijia Wang, Peng Cui, and Shuai Mou.
\newblock Dynamic heterogeneous information network embedding with meta-path based proximity.
\newblock {\em IEEE Transactions on Knowledge and Data Engineering}, 34(3):1117--1132, 2022.

\bibitem[\protect\citeauthoryear{Wang \bgroup \em et al.\egroup }{2023}]{9780569}
Xiao Wang, Deyu Bo, Chuan Shi, Shaohua Fan, Yanfang Ye, and Philip~S. Yu.
\newblock A survey on heterogeneous graph embedding: Methods, techniques, applications and sources.
\newblock {\em IEEE Transactions on Big Data}, 9(2):415--436, 2023.

\bibitem[\protect\citeauthoryear{Xia \bgroup \em et al.\egroup }{2021}]{9416834}
Feng Xia, Ke~Sun, Shuo Yu, Abdul Aziz, Liangtian Wan, Shirui Pan, and Huan Liu.
\newblock Graph learning: A survey.
\newblock {\em IEEE Transactions on Artificial Intelligence}, 2(2):109--127, 2021.

\bibitem[\protect\citeauthoryear{Xue \bgroup \em et al.\egroup }{2020}]{10.1007/978-3-030-67658-2_17}
Hansheng Xue, Luwei Yang, Wen Jiang, Yi~Wei, Yi~Hu, and Yu~Lin.
\newblock Modeling dynamic heterogeneous network for link prediction using hierarchical attention with temporal rnn.
\newblock In {\em Machine Learning and Knowledge Discovery in Databases: European Conference}, page 282–298, 2020.

\bibitem[\protect\citeauthoryear{Yadati \bgroup \em et al.\egroup }{2019}]{10.5555/3454287.3454422}
Naganand Yadati, Madhav Nimishakavi, Prateek Yadav, Vikram Nitin, Anand Louis, and Partha Talukdar.
\newblock Hypergcn: a new method of training graph convolutional networks on hypergraphs.
\newblock In {\em NeurIPS}, 2019.

\bibitem[\protect\citeauthoryear{Yan \bgroup \em et al.\egroup }{2023}]{doi:10.1137/1.9781611977653.ch80}
Bo~Yan, Cheng Yang, Chuan Shi, Jiawei Liu, and Xiaochen Wang.
\newblock Abnormal event detection via hypergraph contrastive learning.
\newblock In {\em Proceedings of the 2023 SIAM International Conference on Data Mining (SDM)}, pages 712--720, 2023.

\bibitem[\protect\citeauthoryear{Yang \bgroup \em et al.\egroup }{2022}]{9300240}
Carl Yang, Yuxin Xiao, Yu~Zhang, Yizhou Sun, and Jiawei Han.
\newblock Heterogeneous network representation learning: A unified framework with survey and benchmark.
\newblock {\em IEEE Transactions on Knowledge and Data Engineering}, 34(10):4854--4873, 2022.

\bibitem[\protect\citeauthoryear{Yang \bgroup \em et al.\egroup }{2023}]{Yang_Yan_Pan_Ye_Fan_2023}
Xiaocheng Yang, Mingyu Yan, Shirui Pan, Xiaochun Ye, and Dongrui Fan.
\newblock Simple and efficient heterogeneous graph neural network.
\newblock {\em AAAI}, 37(9):10816--10824, 2023.

\bibitem[\protect\citeauthoryear{Yu \bgroup \em et al.\egroup }{2017}]{ijcai2017p467}
Wenchao Yu, Wei Cheng, Charu~C Aggarwal, Haifeng Chen, and Wei Wang.
\newblock Link prediction with spatial and temporal consistency in dynamic networks.
\newblock In {\em IJCAI}, pages 3343--3349, 2017.

\bibitem[\protect\citeauthoryear{Zhang \bgroup \em et al.\egroup }{2023a}]{ZHANG2023109486}
Guolin Zhang, Zehui Hu, Guoqiu Wen, Junbo Ma, and Xiaofeng Zhu.
\newblock Dynamic graph convolutional networks by semi-supervised contrastive learning.
\newblock {\em Pattern Recognition}, 139:109486, 2023.

\bibitem[\protect\citeauthoryear{Zhang \bgroup \em et al.\egroup }{2023b}]{9714053}
Mengqi Zhang, Shu Wu, Xueli Yu, Qiang Liu, and Liang Wang.
\newblock Dynamic graph neural networks for sequential recommendation.
\newblock {\em IEEE Transactions on Knowledge and Data Engineering}, 35(5):4741--4753, 2023.

\bibitem[\protect\citeauthoryear{Zhang \bgroup \em et al.\egroup }{2023c}]{10.1145/3527662}
Yingying Zhang, Xian Wu, Quan Fang, Shengsheng Qian, and Changsheng Xu.
\newblock Knowledge-enhanced attributed multi-task learning for medicine recommendation.
\newblock {\em ACM Transactions on Information Systems}, 41(1), 2023.

\bibitem[\protect\citeauthoryear{Zhou \bgroup \em et al.\egroup }{2006}]{NIPS2006_dff8e9c2}
Dengyong Zhou, Jiayuan Huang, and Bernhard Sch\"{o}lkopf.
\newblock Learning with hypergraphs: Clustering, classification, and embedding.
\newblock In {\em NeurIPS}, volume~19, 2006.

\end{thebibliography}

\clearpage

\appendix

\section{Datasets and Baselines}
\label{sec:datasets}

To evaluate the effectiveness of the proposed HTHGN, we conduct experiments on three real-world HTG datasets. The statistics of the dataset are summarized in Table~\ref{tab:dataset}.
\begin{itemize}
    \item \textbf{DBLP}~\footnote{https://dblp.uni-trier.de} is a computer science bibliography website that provides open bibliographic information on major computer science journals and proceedings. The dataset used in the paper consists of 12 network snapshots that contain 8,470 authors, 9,025 papers, and 1,074 venues.
    
    \item \textbf{AMiner}~\footnote{https://www.aminer.org/citation} is an academic search engine that helps us to mine information from academic networks. The paper used a subset of AMiner which contains 12 network snapshots with 8,882 authors, 7,289 papers, and 1,970 venues.
    
    \item \textbf{Yelp}~\footnote{https://www.yelp.com/dataset} is a popular crowd-sourced review platform that allows users to rate and review local businesses. Here we extract a subset of Yelp that is divided into 10 network snapshots by year and consists of 771 businesses, 1,452 users, and 5 stars.
\end{itemize}

\begin{table}[htb]
\centering
\resizebox{.9\linewidth}{!}{
\begin{tabular}{cccc}
\hline
Dataset & Avg. \# Nodes & Avg. \# Edges & \# Snapshots \\ \hline
DBLP &
  \begin{tabular}[c]{@{}c@{}}Author (A): 8,470\\ Paper (P): 9,025\\ Venue (V): 1,074\end{tabular} &
  \begin{tabular}[c]{@{}c@{}}A-P: 8,056\\ A-V: 8,056\\ P-V: 1,903\end{tabular} &
  12 \\
  \hline
AMiner &
  \begin{tabular}[c]{@{}c@{}}Author (A): 8,882\\ Paper (P): 7,289\\ Venue (V): 1,970\end{tabular} &
  \begin{tabular}[c]{@{}c@{}}A-P: 7,538\\ A-V: 7,538\\ P-V: 1,634\end{tabular} &
  12 \\
  \hline
Yelp &
  \begin{tabular}[c]{@{}c@{}}Business (B): 771\\ User (U): 1,452\\ Star (S): 5\end{tabular} &
  \begin{tabular}[c]{@{}c@{}}B-S: 1,080\\ B-U: 1,080\\ U-S: 1,080\end{tabular} &
  9 \\ \hline
\end{tabular}}
\captionof{table}{The Statistics of the Datasets}
\label{tab:dataset}
\end{table}

To comprehensively evaluate the performance of the proposed HTGC model, we compare it with the following homogeneous and heterogeneous GNNs baseline models, and their support for modeling heterogeneity and dynamics is summarized in Table~\ref{tab:baseline}:

\begin{itemize}
    \item \textbf{VGAE} is an unsupervised framework designed for homogeneous static graphs. By optimizing the variational lower bound to model the distribution of hidden states, VGAE achieves competitive results on the static link detection task.
    
    \item \textbf{GATv2} is a more expressive Graph Attention Network (GAT) architecture that overcomes the limitation of GAT. GATv2 aggregates neighbor representations through a modified attention mechanism.

    \item \textbf{DGI} is a cross-view contrastive learning method, which maximizes the mutual information between the local representation of the node and the global representation of the original graph and the row-shuffled negative graph through contrastive learning.
    
    \item \textbf{EvolveGCN} is a homogeneous dynamic graph model that captures the dynamism of the graphs by using a recurrent network to evolve the GCN parameters.

    \item \textbf{DySAT} is a model for capturing the structural evolution of homogeneous dynamic graphs, which captures structural and evolutionary information by a structural attention mechanism on different snapshots and a position-aware temporal self-attention on multiple snapshot representations.

    \item \textbf{HyperGCN} is a hypergraph neural network based on graph spectral theory and propose the hypergraph Laplacian graph convolution for modeling homogeneous hypergraphs. It learns hypergraph structural information by self-supervised training of traditional GCN models.

    \item \textbf{UniGNN} is a unified framework for generalizing general homogeneous GNN models to hypergraphs, which defines four hypergraph message passing mechanisms for classic homogeneous GNNs. We selected the UniGCN and UniGAT models with the best performance among them to conduct comparative experiments.

    \item \textbf{HGNNP} is a hypergraph neural network framework for modeling homogeneous hypergraphs, which completes two-step message-passing by performing general spatial convolution on the hypergraph.
    
    \item \textbf{metapath2vec} is a scalable heterogeneous static graph representation learning model that uses metapath-based random walks and skip-gram models to model heterogeneous structural and semantic information in HG.
    
    \item \textbf{R-GCN} is proposed to model knowledge graphs that contain multiple relationships. R-GCN preserves the heterogeneity in the graph structure by separately transforming and aggregating neighbor nodes with different relations.

    \item \textbf{HGT} is an architecture to model heterogeneous static graph. In order to capture graph heterogeneity, HGT uses the triplet meta-relation for different relations to parameterize the attention weight matrices.

    \item \textbf{H-GVAE} is the static HGs conditional variational autoencoder model HetSANN-GVAE, which models the latent probability distribution of heterogeneous nodes through an attention mechanism on heterogeneous graphs.

    \item \textbf{HPN} is a static heterogeneous graph model used to alleviate semantic confusion issues, which improves node-level aggregation through the semantic propagation mechanism, and obtains rich semantic representation through the semantic fusion mechanism.

    \item \textbf{DyHATR} is a model designed for HTGs, which models the heterogeneity and dynamics of different snapshots in HTGs by heterogeneous hierarchical attention and recurrent neural network, and finally aggregates different snapshot representations by temporal self-attention mechanism.

    \item \textbf{HTGNN} is a model for modeling dynamics and heterogeneity in HTGs by applying a hierarchical attention mechanism on different snapshots respectively and aggregating the dynamics among different snapshots through a temporal attention mechanism.
    
\end{itemize}

\begin{table}[htb]
\centering
\small
\begin{tabular}{cccc}
\hline
Methods      & Hete. & Temp. & Hyper. \\ \hline
VGAE~\cite{kipf2016variational}         & \xmark    & \xmark  & \xmark  \\
GATv2~\cite{brody2022how}          & \xmark    & \xmark  & \xmark  \\
DGI~\cite{veličković2018deep}          & \xmark    & \xmark  & \xmark  \\
EvolveGCN~\cite{EvolveGCN2020}    & \xmark    & \cmark  & \xmark  \\
DySAT~\cite{SankarDySAT}        & \xmark    & \cmark  & \xmark  \\
metapath2vec~\cite{10.1145/3097983.3098036} & \cmark    & \xmark  & \xmark  \\
HyperGCN~\cite{10.5555/3454287.3454422}    & \xmark    & \xmark  & \cmark  \\
UniGNN~\cite{ijcai2021p353}      & \xmark    & \xmark  & \cmark  \\
HGNNP~\cite{9795251}       & \xmark    & \xmark  & \cmark  \\
R-GCN~\cite{SchlichtkrullRGCN}        & \cmark    & \xmark  & \xmark  \\
HGT~\cite{10.1145/3366423.3380027}          & \cmark    & \xmark  & \xmark  \\
HetSANN-GVAE~\cite{dalvi2022variational}    & \cmark    & \xmark  & \xmark  \\
HPN~\cite{9428609}          & \cmark    & \xmark  & \xmark  \\
DyHATR~\cite{10.1007/978-3-030-67658-2_17}       & \cmark    & \cmark  & \xmark  \\
HTGNN~\cite{doi:10.1137/1.9781611977172.74}        & \cmark    & \cmark  & \xmark  \\
HTHGN         & \cmark    & \cmark   & \cmark \\ 
\hline
\end{tabular}
\captionof{table}{Comparison of Baseline Methods in Supporting Heterogeneity (Hete.), Temporal Dynamics (Temp.), and Hyper-Relational Modeling (Hyper.)}
\label{tab:baseline}
\end{table}

\begin{table*}[!htb]
\centering
\begin{tabular}{ccccccc}
\hline
Dataset & \multicolumn{2}{c}{Yelp}                        & \multicolumn{2}{c}{DBLP}                        & \multicolumn{2}{c}{AMiner}                      \\ \hline
Metrics & AUC                    & AP                     & AUC                    & AP                     & AUC                    & AP                     \\
1-hop   & 70.97 $\pm$ 5.86       & 68.96 $\pm$ 7.32       & 88.66 $\pm$ 2.43       & 89.24 $\pm$ 2.34       & 95.39 $\pm$ 1.66       & 95.56 $\pm$ 1.59       \\
2-hop   & 72.33 $\pm$ 3.30       & 70.27 $\pm$ 4.83       & 89.39 $\pm$ 2.39       & 89.77 $\pm$ 2.23       & 95.36 $\pm$ 1.67       & 95.50 $\pm$ 1.60       \\
3-hop   & 72.17 $\pm$ 3.22       & {\ul 70.36 $\pm$ 4.54} & 89.28 $\pm$ 2.40       & 89.73 $\pm$ 2.25       & {\ul 95.51 $\pm$ 1.76} & {\ul 95.62 $\pm$ 1.66} \\ \hline
1-ring  & 70.71 $\pm$ 5.63       & 68.70 $\pm$ 7.08       & 88.56 $\pm$ 2.51       & 89.18 $\pm$ 2.46       & 95.28 $\pm$ 1.78       & 95.46 $\pm$ 1.68       \\
2-ring  & {\ul 72.35 $\pm$ 3.24} & 70.32 $\pm$ 4.72       & {\ul 89.48 $\pm$ 2.59} & {\ul 89.88 $\pm$ 2.44} & 95.50 $\pm$ 1.66       & 95.59 $\pm$ 1.59       \\
3-ring &
  \textbf{74.04 $\pm$ 4.82} &
  \textbf{89.56 $\pm$ 3.01} &
  \textbf{91.33 $\pm$ 1.61} &
  \textbf{96.97 $\pm$ 0.56} &
  \textbf{96.58 $\pm$ 1.14} &
  \textbf{98.80 $\pm$ 0.40} \\ \hline
\end{tabular}%
\caption{AUC and AP scores for HTHGN with different hyperedge types in link prediction on three datasets.}
\label{tab:klinkpred}
\end{table*}

\begin{table*}[!htb]
\centering
\begin{tabular}{ccccccc}
\hline
Dataset & \multicolumn{2}{c}{Yelp}                              & \multicolumn{2}{c}{DBLP}                        & \multicolumn{2}{c}{AMiner}                      \\ \hline
Metrics & AUC                       & AP                        & AUC                    & AP                     & AUC                    & AP                     \\
1-hop   & 62.27 $\pm$ 7.23          & 63.89 $\pm$ 9.60          & 76.65 $\pm$ 2.01       & {\ul 76.09 $\pm$ 1.09} & 89.76 $\pm$ 1.59       & 90.13 $\pm$ 1.15       \\
2-hop   & \textbf{63.20 $\pm$ 3.66} & {\ul 65.47 $\pm$ 6.37}    & 77.04 $\pm$ 2.48       & 74.83 $\pm$ 2.32       & 89.70 $\pm$ 1.47       & 89.86 $\pm$ 1.22       \\
3-hop   & {\ul 62.90 $\pm$ 3.63}    & \textbf{65.55 $\pm$ 6.23} & 76.45 $\pm$ 2.39       & 74.32 $\pm$ 2.21       & 89.96 $\pm$ 1.63       & 90.03 $\pm$ 1.22       \\ \hline
1-ring  & 61.75 $\pm$ 6.86          & 63.26 $\pm$ 9.12          & 76.29 $\pm$ 2.18       & 75.88 $\pm$ 1.49       & 89.49 $\pm$ 1.69       & 89.83 $\pm$ 1.40       \\
2-ring  & 62.82 $\pm$ 3.83          & 65.19 $\pm$ 6.39          & {\ul 77.36 $\pm$ 2.22} & 75.17 $\pm$ 1.82       & {\ul 90.09 $\pm$ 1.52} & {\ul 90.17 $\pm$ 1.29} \\
3-ring & 61.39 $\pm$ 4.35 & 63.99 $\pm$ 6.55 & \textbf{81.77 $\pm$ 1.66} & {\ul \textbf{79.43 $\pm$ 1.88}} & \textbf{91.93 $\pm$ 1.02} & \textbf{91.04 $\pm$ 1.25} \\ \hline
\end{tabular}%
\caption{AUC and AP scores of HTHGN with different hyperedge types in new link prediction on three datasets.}
\label{tab:knewlinkpred}
\end{table*}

\section{Experiment Setup Details}
\label{app:expsetup}

We conducted dynamic link prediction experiments to verify the gain of higher-order interactions on representation learning performance. In link prediction and new link prediction experiments, we evaluate the performance of HTHGN and baselines in classifying true and false edges between heterogeneous nodes. Specifically, we held out the last $3$ snapshots for testing and trained the model on the remaining snapshots. We use $T$ snapshots before the test snapshot as input in each evaluation and generate node representations. Then, take the average score on the last $3$ snapshots as the result of an experiment. 
In link prediction of the dynamic graph learning context, the difference between link prediction and new link prediction, as stated in Definition~\ref{def:linkpred}, is due to the different selection of true edges for evaluation. Specifically, link prediction uses all heterogeneous edges in future moments as positive edges, while new link prediction only uses edges that have not appeared among them as positive edges. We sample an equal number of false edges for the link prediction task to evaluate the prediction performance. We repeat this process and calculate the average AUC and AP scores on the last $3$ snapshots as the prediction results. In the new link prediction, we first identify never-before-seen edges as true edges and evaluate the model under a similar setup to the link prediction experiment. Compared with link prediction, new link prediction is more challenging because it focuses on edges that have not yet appeared. We performed $5$ repeated randomized experiments for all methods and reported their means and standard deviations. 

\section{Implementation Details}
\label{sec:implementation}
For the proposed model HTHGN, we use Glorot initialization and optimize the model with Adam optimizer.
We set the initial learning rate of the Adam optimizer to 1e-3 and the regularization parameter to 5e-4. 
The dropout rate for attention is set to $0.2$.
We set $k=3$ and $p=100$ by grid search and train all the models with $300$ epochs.
We conduct a grid search in the embedding dimension from $4$ to $64$ and set it to $32$ for each dataset and baseline method for fair comparison.
All models are randomly trained for $5$ times, and we report the average test performance results.

Since HTGs cannot be directly applied to the homogeneous and static graph baselines, for the homogeneous graph method, we masked the types of nodes and relationships to verify the effectiveness of heterogeneous information. For the static graph method, we merge the training graph snapshots to remove the temporal evolution. For the random walk method metapath2vec based on meta-path guidance, we set the number of steps per node to 40, the step size to 60, and the window size in the skip-gram to 5. For its meta-path, we set \textit{\{APVPA, PVAVP, VAPAV\}} for the academic network DBLP and AMiner datasets, and \textit{\{BSUSB, SUBUS, UBSBU\}} for the Yelp dataset. For the hypergraph learning method, we uniformly use $3$-hop neighbors through hyperedges to construct group interaction relationships based on the network structure and remove the type information of nodes to adapt to the model. For the rest of the models, we followed the parameter setting of the best performance reported in the paper. All models are randomly trained for $5$ times, and we report the average test performance results.

Our experiments were conducted on the Ubuntu 18.04.6 LTS operating system with Intel(R) Xeon(R) Silver 4210R CPU @ 2.40GHz processor and NVIDIA GeForce RTX 3090 GPUs. The CUDA version uses 11.8, and the PyTorch version uses 2.0.1. We implement the proposed model using Deep Graph Library (DGL) 1.1.1.

\section{Complexity Analysis}
\label{sec:complexity}
Here we conduct a detailed analysis of each part of HTHGN and its overall complexity, which is mainly composed of Heterogeneous and Temporal Attention Aggregation modules. During the implementation of HTHGN, parallel strategies such as parallelizing across each snapshot on the temporal receptive field $T$ are used for efficient inference. 
Computationally, it is highly efficient: the feature projection complexity is $\mathcal{O}\left(|V_*|Dd\right)$.
The complexity of heterogeneous attention aggregation that can be parallelized over $T$ snapshots is $\mathcal{O}\left((|V_*|+|E_*|)d\right)$.
The temporal attention aggregation complexity is $\mathcal{O}\left(|V_*|dT^2\right)$.
The complexity of the contrastive loss functions is $\mathcal{O}\left(|E_*|Qd\right)$.
Overall, the complexity of HTHGN is $\mathcal{O}\left(|V_*|Dd + |V_*|dT^2 + |E_*|d\right)$, which is efficient in a linear relationship with the node and edge set cardinality.

\section{Additional Experiment Results}
\label{app:exps}

We report experimental results not presented in the main text due to space limitations.

\subsection{New Link Prediction}
\label{app:newlp}

We also evaluate the performance of the new link prediction on three datasets against the proposed HTHGN and baselines. As shown in Table~\ref{tab:newlinkpred}, the proposed HTHGN has obvious performance advantages on all three datasets and achieves the best results in both its AUC and AP scores. It is worth noting that because new link predictions are only evaluated on edges that have not appeared before, the dynamic graph model achieves better results than link prediction. In addition, homogeneous graph methods, whether low-order or high-order GNN, are generally worse than heterogeneous graph representation learning methods due to their inability to model different types of semantic information. We would like to point out that DyHATR and HTGNN generally surpass the other baselines due to their novel designs that simultaneously capture dynamics and heterogeneity, which shows that temporal evolution and heterogeneous semantic information are critical to representation learning. Nonetheless, our proposed HTHGN can simultaneously capture high-order heterogeneous evolutionary relationships and achieve significant performance improvements, which verifies the effectiveness of modeling group relationships.

\begin{table*}[!htb]
\resizebox{\textwidth}{!}{%
\begin{tabular}{cccccccc}
\hline
Dataset      & \multicolumn{2}{c}{Yelp}            & \multicolumn{2}{c}{DBLP}            & \multicolumn{2}{c}{AMiner}          & \multirow{2}{*}{Avg. Rank} \\ \cline{1-7}
Metrics      & AUC              & AP               & AUC              & AP               & AUC              & AP               &                            \\ \hline
VGAE         & 45.71 $\pm$ 3.69 & 51.68 $\pm$ 4.38 & 51.88 $\pm$ 2.19 & 58.30 $\pm$ 1.58 & 61.26 $\pm$ 3.23 & 69.12 $\pm$ 2.51 & 9.67                       \\
GATv2        & 46.21 $\pm$ 1.29 & 48.17 $\pm$ 1.22 & 63.97 $\pm$ 0.65 & 64.74 $\pm$ 0.47 & 70.30 $\pm$ 0.65 & 73.15 $\pm$ 0.92 & 6.50                       \\
DGI          & 46.84 $\pm$ 4.21 & 51.99 $\pm$ 4.55 & 49.38 $\pm$ 2.48 & 55.11 $\pm$ 0.85 & 58.70 $\pm$ 4.05 & 65.62 $\pm$ 3.42 & 10.83                      \\
EvolveGCN    & 50.48 $\pm$ 5.21 & 51.80 $\pm$ 4.30 & 55.92 $\pm$ 3.28 & 57.24 $\pm$ 2.34 & 60.50 $\pm$ 4.88 & 64.05 $\pm$ 4.14 & 9.00                       \\
DySAT        & 49.19 $\pm$ 1.63 & 51.10 $\pm$ 1.22 & 58.06 $\pm$ 1.42 & 59.61 $\pm$ 1.35 & 63.18 $\pm$ 0.95 & 65.86 $\pm$ 1.15 & 8.00                       \\
HyperGCN     & 50.00 $\pm$ 1.19 & 50.80 $\pm$ 1.67 & 59.28 $\pm$ 0.83 & 59.45 $\pm$ 0.96 & 60.68 $\pm$ 0.46 & 61.20 $\pm$ 0.98 & 9.33                       \\
UniGCN       & 44.53 $\pm$ 0.78 & 47.79 $\pm$ 0.85 & 46.07 $\pm$ 2.18 & 47.33 $\pm$ 1.99 & 52.39 $\pm$ 1.82 & 52.94 $\pm$ 2.10 & 16.33                      \\
UniGAT       & 46.07 $\pm$ 1.81 & 49.25 $\pm$ 1.90 & 56.28 $\pm$ 0.98 & 53.76 $\pm$ 0.95 & 58.73 $\pm$ 1.83 & 59.59 $\pm$ 1.83 & 12.50                      \\
HGNNP        & 47.25 $\pm$ 1.72 & 50.23 $\pm$ 1.66 & 51.55 $\pm$ 0.43 & 50.50 $\pm$ 0.33 & 52.72 $\pm$ 1.86 & 54.67 $\pm$ 0.94 & 14.00                      \\ \hline
metapath2vec & 50.88 $\pm$ 0.72 & 51.22 $\pm$ 0.64 & 52.75 $\pm$ 1.32 & 51.78 $\pm$ 1.01 & 57.45 $\pm$ 2.14 & 56.34 $\pm$ 1.69 & 11.50                      \\
R-GCN        & 52.26 $\pm$ 1.29 & 51.37 $\pm$ 0.81 & 59.73 $\pm$ 6.98 & 61.08 $\pm$ 7.46 & 69.03 $\pm$ 3.38 & 72.15 $\pm$ 3.48 & 5.17                       \\
HGT          & 48.88 $\pm$ 2.07 & 49.93 $\pm$ 2.13 & 57.94 $\pm$ 0.91 & 59.80 $\pm$ 1.19 & 60.78 $\pm$ 1.27 & 66.31 $\pm$ 1.28 & 9.00                       \\
HetSANN-GVAE & 38.49 $\pm$ 1.57 & 43.45 $\pm$ 1.21 & 52.36 $\pm$ 1.23 & 53.11 $\pm$ 1.54 & 58.31 $\pm$ 1.84 & 63.11 $\pm$ 1.29 & 14.50                      \\
HPN          & 43.04 $\pm$ 2.36 & 46.70 $\pm$ 1.99 & 62.12 $\pm$ 1.77 & 61.42 $\pm$ 2.00 & 63.15 $\pm$ 1.59 & 69.54 $\pm$ 1.20 & 9.17                       \\
DyHATR &
  {\ul 53.39 $\pm$ 0.92} &
  {\ul 55.33 $\pm$ 1.41} &
  63.30 $\pm$ 1.40 &
  62.20 $\pm$ 1.18 &
  72.44 $\pm$ 1.47 &
  70.98 $\pm$ 1.75 &
  3.33 \\
HTGNN &
  52.44 $\pm$ 1.11 &
  51.14 $\pm$ 1.80 &
  {\ul 70.62 $\pm$ 1.84} &
  {\ul 69.45 $\pm$ 1.66} &
  {\ul 75.53 $\pm$ 2.47} &
  {\ul 74.15 $\pm$ 1.61} &
  {\ul 3.17} \\ \hline
HTHGN &
  \textbf{64.93 $\pm$ 1.91} &
  \textbf{68.47 $\pm$ 1.42} &
  \textbf{82.78 $\pm$ 1.36} &
  \textbf{78.28 $\pm$ 1.08} &
  \textbf{93.01 $\pm$ 0.78} &
  \textbf{91.91 $\pm$ 1.26} &
  \textbf{1.00} \\ \hline
\end{tabular}%
}
\caption{AUC and AP scores of new link prediction tasks between HTHGN and baselines in three datasets.}
\label{tab:newlinkpred}
\end{table*}

\subsection{Impact of Hypergraph Construction}
\label{app:param-hyper}

Here we report the performance of HTHGN on new link prediction tasks in three real-world datasets under different $k$ values and different hyperedge types. As shown in Table~\ref{tab:knewlinkpred}, HTHGN's new link prediction performance under different configurations is generally stable. Moreover, as $k$ in the $k$-ring hyperedge increases, the expansion of the receptive field generally leads to better model performance on multiple datasets. However, there is a certain performance degradation in $k$-hop hyperedges. We believe this is because the low-order pairwise relationships already contain 2-hop intermediate node information. Repeatedly introducing them through hyperedges is not conducive to modeling high-order groups. interaction. Finally, similar to the link prediction task, $3$-ring hypergraphs still generally achieve the best performance in new link prediction, which verifies our motivation that group interaction is beneficial to HTG representation learning.

\subsection{Impact of $P$-uniform}
\label{app:puniform}

\begin{figure*}[!htb]
    \begin{minipage}[b]{0.49\linewidth}
    \centering
    \begin{subfigure}[b]{0.49\linewidth}
        \includegraphics[width=\linewidth]{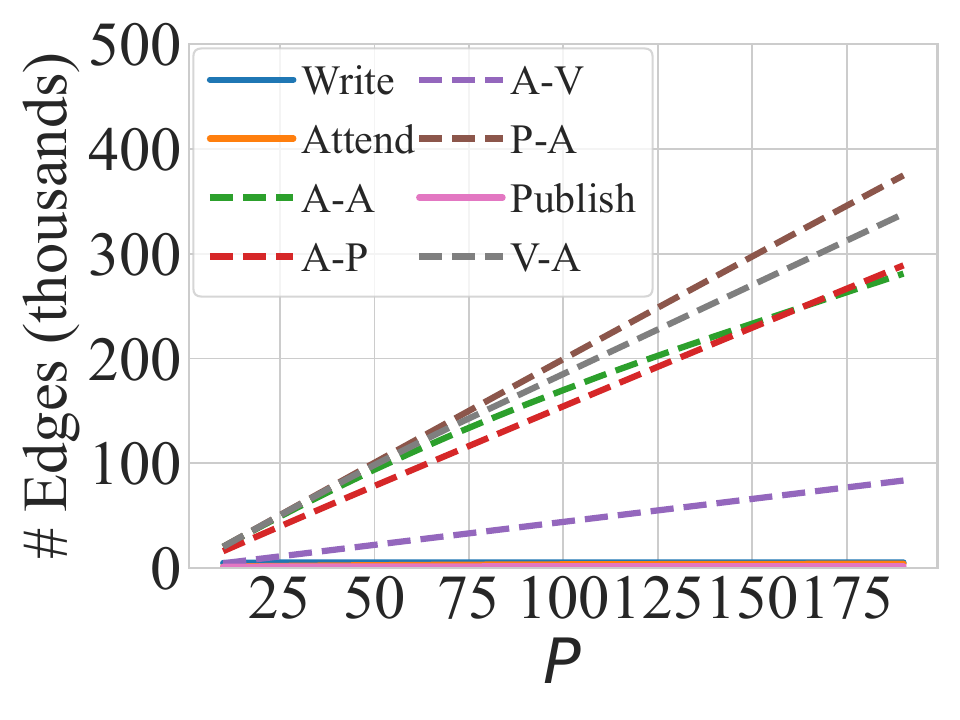}
        \caption{AMiner Dataset}
    \end{subfigure}
    \hfill
    \begin{subfigure}[b]{0.49\linewidth}
        \includegraphics[width=\linewidth]{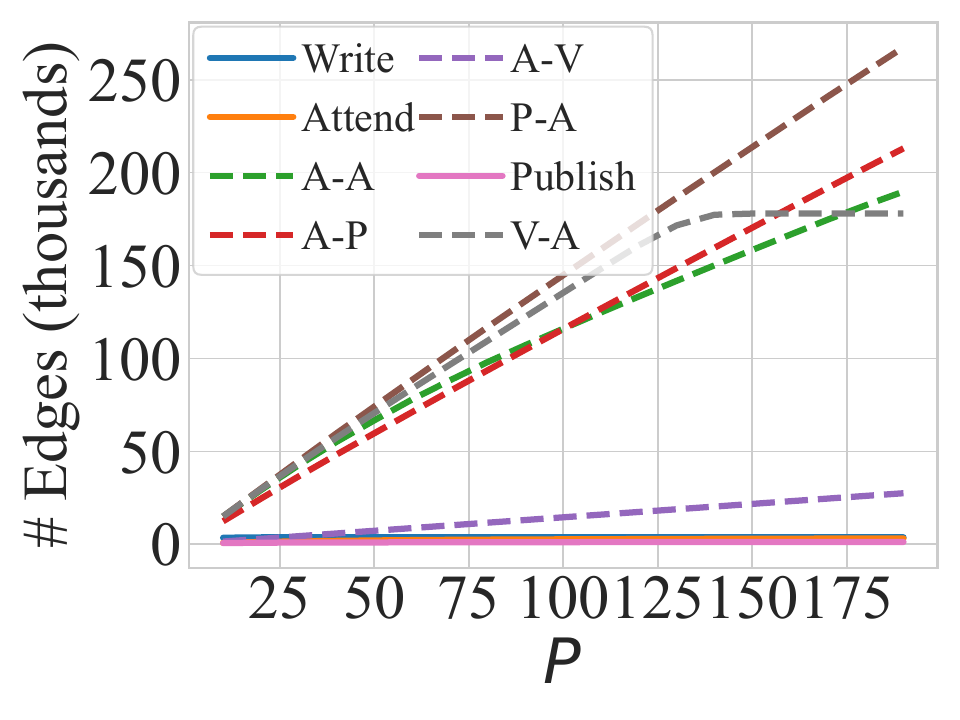}
        \caption{DBLP Dataset}
    \end{subfigure}
    \hfill
    \caption{Number of edges of $P$-uniform on AMiner and DBLP datasets.}
    \label{fig:uniform-app}    
    \end{minipage}
    \hfill
    \begin{minipage}[b]{0.49\linewidth}
    \centering
    \begin{subfigure}[b]{\linewidth}
        \includegraphics[width=\linewidth]{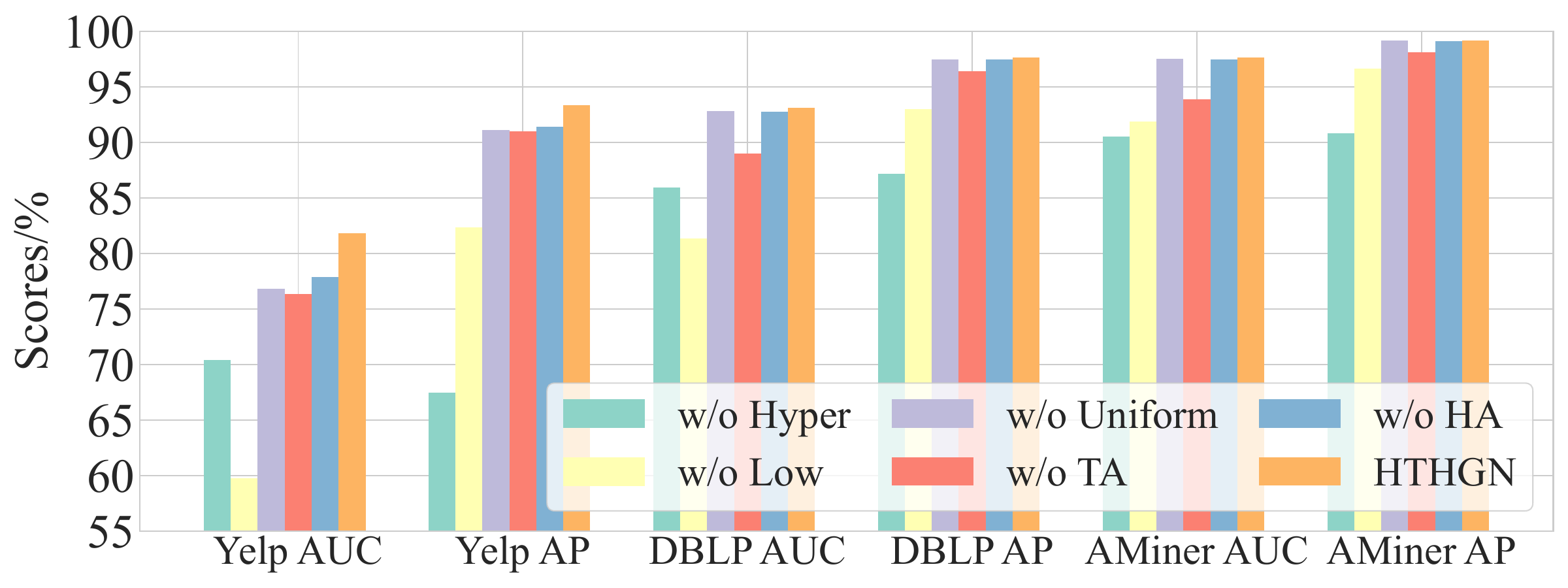}
    \end{subfigure}
    \caption{Ablation experiments of link prediction tasks on three datasets.}
    \label{fig:lpablation}
    \end{minipage}
\end{figure*}

We tested the number of $P$-uniform hyperedges and low-order edges in the AMiner and DBLP datasets under different $P$ values, and the results are reported in Figure~\ref{fig:uniform-app}. The visualization shows that as $P$ increases, the number of hyperedges increases. Therefore, this result shows that compared with the naive cluster/star expansion algorithm, HTHGN can significantly improve computational efficiency by controlling the $P$-uniform hyperedges.

\subsection{Ablation Study}
\label{app:ablation}
We conducted ablation experiments on various main parts of the proposed HTHGN method to verify the effectiveness of the module. The experimental results of link prediction in three data sets are shown in Figure~\ref{fig:lpablation}. Similar to the analysis results in the main text, each part plays a non-negligible role in modeling higher-order relationships, which shows the effectiveness of each module of HTHGN.

\end{document}